\documentclass[showpacs,prb,amssymb]{revtex4}
\usepackage{graphicx}
\usepackage[tbtags]{amsmath}
\usepackage{bm}
\usepackage[tbtags]{amsmath}

\begin{document}

\title{The Voltage-Current Characteristic of high $T_{C}$ DC SQUID:
theory, simulation, experiment}
\author{Ya. S. Greenberg, I. L. Novikov}
\affiliation{Novosibirsk State Technical University, 20 K. Marx
Ave., 630092 Novosibirsk, Russia}

\date{\today}

\begin{abstract}
The analytical theory for the voltage-current characteristics of
the large inductance ($L>100$pH) high-$T_C$ DC SQUIDs that has
been developed previously is consistently compared with the
computer simulations and the experiment. The theoretical voltage
modulation for symmetric junctions is shown to be in a good
agreement with the results of known computer simulations. It is
shown that the asymmetry of the junctions results in the increase
of the voltage modulation if the critical current is in excess of
some threshold value (about $8\mu A$). Below this value the
asymmetry leads to the reduction of the voltage modulation as
compared to the symmetric case. The comparison with the experiment
shows that the asymmetry can explain a large portion of
experimental values of the voltage modulation which lie above the
theoretical curve for symmetric DC SQUID. It also explains
experimental points which lie below the curve at small critical
currents. However, a significant portion of these values which lie
below the curve cannot be explained by the junction asymmetry.
\end{abstract}
\maketitle

\section{Introduction}
The new type of superconductors, which has been discovered in the
end of the last century by Bednorz and Muller, is widely used in
the modern SQUID systems. The majority of high $T_C$ DC SQUIDs are
based on the YBCO thin films and have the different types of
design \cite{Beyer, Park, Jia}. However, the adequate theory of
the voltage-current characteristic (VCC) of the high-$T_C$ DC
SQUID, which would predict its transfer function and energy
resolution, still not exists. In recent time intensive computer
simulations and theoretical studies have been performed to
investigate the dependence of high-$T_C$ DC SQUID behavior on
various factors \cite{Enpuku, Chesca, Koelle}, but a marked
disagreement of the numerical simulations with experiment is still
observed: the experimental transfer functions in many cases are
much lower than the values predicted by theory and computer
simulations; the white noise is about ten times higher than
predicted. This is one of the most important unsolved problems,
which seriously hinders the optimization of high $T_C$ DC SQUIDs
for applications.

Up to now the high-$T_C$ DC SQUIDs have the significant parameter
dispersion, but the reasons of such dispersion are not
established. One of the possible reasons for the dispersion could
be attributed to the junction asymmetry of SQUID interferometer
(unequal critical currents or (and) normal resistances), which for
grain boundary junctions is about 20-30 percents due to on chip
technological heterogeneity.

Recently, the theory of the voltage-current characteristic of the high $T_C$ DC SQUID, which expands the validity range of the Chesca's analytic
theory \cite{Chesca} on the DC SQUIDs with large inductance $L>100$ pH and accounts both for the symmetric and asymmetric DC SQUIDs, was
developed \cite{Green1, Green2}. The theory is based on the perturbation solution of the two-dimensional Fokker-Planck equation (2D FPE) that
describes the stochastic dynamics of DC SQUID in presence of the large thermal fluctuations.

In the paper the results of the analytic theory of VCC of high
$T_C$ DC SQUID \cite{Green1, Green2} are compared with the
computer simulations and with the experiment. It is shown that the
theoretical voltage modulation for symmetric junctions is in a
good agreement with the results of known computer simulations. It
is also shown that the asymmetry of the junctions results in the
increase of the voltage modulation if the critical current is in
excess of some threshold value (about $8\mu A$). Below this value
the asymmetry leads to the reduction of the voltage modulation as
compared to the symmetric case. The comparison with the experiment
shows that the asymmetry can explain a large portion of
experimental values of the voltage modulation which lie above the
theoretical curve for symmetric DC SQUID. It also explains
experimental points which lie below the curve at small bias
currents (less about than $10 \mu A$). However, a significant
portion of these values which lie below the curve at larger bias
currents cannot be explained by the junction asymmetry.

The paper is organized as follows. In the section II we present in
a concise form the main theoretical results for symmetric and
asymmetric DC SQUIDs. In Section III we compare theoretical
voltage-current characteristics for symmetric DC SQUID with the
computer simulations of stochastic dynamical equations of DC
SQUIDs which have been made earlier by other authors. In this
section we also study in detail the influence of the junction
asymmetry on the voltage modulation as compared with symmetric
case. In Section IV we compare the theory with experiment and show
that the junction asymmetry can explain a large portion of
experimental points which lie well above the theoretical voltage
modulation curve for symmetric DC SQUID.

\section{The main results of the analytic theory of VCC of high $T_C$
DC SQUID}
\subsection{Symmetric DC SQUID}
We consider a symmetric DC SQUID with equal critical currents of
junctions, $I_{C1}=I_{C2}\equiv I_C$, equal normal resistance,
$R_1=R_2\equiv R$, and loop inductance $L$. The equations,
describing such SQUID, have the following form:

\begin{equation}\label{flux}
    \frac{L}{{2R}}\frac{{d\Phi }}{{dt}} = \Phi _X  - \Phi  - LI_C
    \cos \delta \sin \varphi  + \frac{L}{2}I_{N_ - } (t)
\end{equation}

\begin{equation}\label{delta}
    \frac{{\Phi _0 }}{{\pi R}}\frac{{d\delta }}{{dt}} = I - 2I_C
    \sin \delta \cos \varphi  - I_{N _ {+}}  (t)
\end{equation}
where $\Phi_0 = h/2e$ is a quantum of magnetic flux, $\Phi$ is a
magnetic flux trapped in the interferometer loop, $\Phi_X$ is
external magnetic flux, $I$ is a bias current,
$\varphi=\pi\Phi/\Phi_0$, $\delta=\pi\Delta/\Phi_0$. The
quantities $I_{N\pm}(t)$ are independent stochastic variables
related to the Nyquist current noise of junctions:
\begin{equation}\label{currnoise}
    \left\langle {I_{N _ \pm}  (t)I_{N _ \pm}  (t')} \right\rangle  =
    \frac{{4k_B T}}{R}\delta (t - t')
\end{equation}
where $k_Â$ is the Boltzmann constant, $T$ is the absolute
temperature.

The output voltage across SQUID is the low frequency part of the
equation (\ref{delta}), which is averaged over the noise:
\begin{equation}\label{volt}
    V = \frac{{\Phi _0 }}{{2\pi }}\left\langle {\frac{{d\delta }}{{dt}}} \right\rangle
\end{equation}

Thus, the stochastic equations (\ref{flux}), (\ref{delta}) are
described DC SQUID behavior in presence of the thermal
fluctuations. The solution of these equations depends on the
following parameters: screening parameter $\beta=2LI_C/\Phi_0$,
noise parameter $\Gamma=2\pi k_BT/\Phi_0I_C$ and dimensionless
inductance $\alpha=L/L_F$, where $L_F = (\Phi_0/2\pi)^2/k_BT$ is a
fluctuation inductance, equaled 100 pH at $Ò$= 77 K. But only two
of them are independent due to relation $\alpha=\pi\beta\Gamma$.

The known theoretical approaches to the solution of equations (\ref{flux}), (\ref{delta}) are based on the analysis of the equivalent two
dimensional Fokker-Planck equation for the distribution function $P(\Phi,\Delta)$:
\begin{equation}\label{FPE}
    \frac{1}{{2R}}\frac{{\partial P}}{{\partial t}} = \frac{\partial }{{\partial
    \Phi }}\left( {\frac{{\partial W}}{{\partial \Phi }}P} \right) +
    \frac{\partial }{{\partial \Delta }}\left( {\frac{{\partial W}}{{\partial
     \Delta }}P} \right) + k_B T\frac{{\partial ^2 P}}{{\partial \Phi ^2 }} +
     k_B T\frac{{\partial ^2 P}}{{\partial \Delta ^2 }}
\end{equation}
where $W$ is a two dimensional potential energy for DC SQUID:
\begin{equation}\label{Energy}
    W = E_J \left[ {1 - \cos \varphi \cos \delta  - \frac{i}{2}\delta  +
    \frac{{\left( {\varphi  - \varphi _X } \right)^2 }}{{\pi \beta }}} \right]
\end{equation}
$E_J=\Phi_0I_C/2\pi$  is a Josephson coupling energy of two
junctions in parallel, $i=I/I_C$, $\phi_X=\pi\Phi_X/\Phi_0$ is a
dimensionless external magnetic flux.

At the present time there exist some analytical solutions of
equation (\ref{FPE}), which are valid at different ranges of
$\beta, \Gamma, \alpha$.

1) The exact analytical solution of FPE can be obtained in the
small inductance limit $(L\rightarrow 0, \beta<< 1, \alpha<<1)$.
In this case SQUID is equivalent to a single Josephson junction
with normal resistance $R/2$ and critical current
$2I_C|\cos\phi_X|$. The voltage across such single junction is
well known and can be obtained from Ambegaokar-Halperin form
\cite{Amb}:
\begin{equation}\label{voltAmb}
    \frac{V}{{RI_C }} = \frac{{\pi \Gamma }}{{p\left( {i,\Gamma ,\varphi _X }
    \right)}}
\end{equation}
where
\begin{equation}\label{prob}
    p\left( {i,\Gamma ,\varphi _X } \right) = \left[ {\int\limits_0^{2\pi }
     {e^{ - W(y)} dy\int\limits_0^y {e^{W(x)} dx}  - \left( {1 - e^{\frac{{2\pi i}}
     {\Gamma }} } \right)^{ - 1} \int\limits_0^{2\pi } {e^{W(x)} dx}
     \int\limits_0^{2\pi } {e^{ - W(x)} dx} } } \right]
\end{equation}
\begin{equation}\label{Pot}
    W(x) = \left( {i/\Gamma } \right)x + \left( {2/\Gamma }
     \right)\cos \varphi _X \cos x
\end{equation}
2) The approximate analytical solution of FPE (\ref{FPE}) was
obtained in \cite{Chesca}. This solution is valid at $\beta< 0.3$
and relatively high thermal fluctuations level ($\Gamma>1$).

3) The original method for the solution of equation (\ref{FPE})
for SQUID with large inductance ($\alpha\geq 1$) was presented in
\cite{Green1}. The method is based on the perturbation expansion
of the solution of 2D FPE (\ref{FPE}) over small parameter
$\varepsilon=\exp(-\alpha/2)$. The result is the following
expressions for voltage V across a SQUID and voltage modulation
$\Delta V=V(\phi_X=\pi/2)-V(\phi_X=0)$:
\begin{equation}\label{voltsym}
    \frac{V}{{RI_C }} = \frac{{2\pi \Gamma }}{{p\left( {i,2\Gamma ,0}
    \right)}} - \frac{1}{2}\exp \left( { - \alpha /2} \right)\cos
    (2\varphi _X )f(i,\Gamma )
\end{equation}
\begin{equation}\label{voltmod}
    \frac{{\Delta V}}{{RI_C }} = \exp \left( { - \alpha /2} \right)f(i,\Gamma )
\end{equation}
where the value $p(i, 2\Gamma,0)$ was defined in (\ref{prob}).

According to \cite{Green2}
\begin{equation}\label{f}
    f(i,\Gamma ) = \frac{{256\pi ^3 i^2 \Gamma ^3 }}{{\left[ {p\left( {i,2\Gamma ,0}
     \right)} \right]^3 }}B^2
\end{equation}
where
\begin{equation}\label{B}
    B = \sum\limits_{n =  - \infty }^{n =  + \infty } {\frac{{( - 1)^n I_n
    \left( {\frac{1}{\Gamma }} \right)I_{n + 1} \left( {\frac{1}{\Gamma }}
    \right)}}{{i^2  + 4n^2 \Gamma ^2 }}}
\end{equation}
$I_n$ is a modified Bessel function.

These expressions are valid for $\alpha\geq1$ and any values of
$\beta$, and $\Gamma$ which are consistent with the condition
$\alpha=\pi\beta\Gamma$. Therefore, they can be applied for the
analysis of a majority of practical high $T_C$ DC SQUIDs with
$\Gamma\approx0.05-1$, $\beta\geq1$, $\alpha\geq1$. However, it
should be remembered that Eqs.(\ref{volt}) and (\ref{voltmod}) are
the approximate expressions which account for the first order term
in the perturbation expansion of the voltage over small parameter
$\varepsilon=\exp(-\alpha/2)$.

\subsection{Asymmetric DC SQUID}

We describe the junction asymmetry in terms of the asymmetry
parameters $\gamma$ and $\rho$, which are defined according to:
$I_{C1}=(1+\gamma)I_C$, $I_{C2}=(1-\gamma)I_C$, $R_1=R/(1+\rho)$,
$R_2=R/(1-\rho)$, where
\begin{eqnarray}\label{critcurr}
    I_C=\frac{I_{C1}+I_{C2}}{2} ;&&\gamma=\frac{I_{C1}-I_{C2}}{I_{C1}+I_{C2}}
\end{eqnarray}
\begin{eqnarray}\label{Resist}
    R = \frac{{2R_1 R_2 }}{{R_1  + R_2 }} ;&&{\rho  = \frac{{R_2  - R_1 }}{{R_1  + R_2 }}} > 0
\end{eqnarray}
The perturbation  method, developed in \cite{Green1}, has been
applied to obtain  the expressions for the voltage and its
modulation across an asymmetric SQUID with large inductance
\cite{Green2}. Corresponding expressions have the following forms:
\begin{equation}\label{VoltAsym}
    \frac{V}{{RI_C }} = \frac{{V_0 }}{{RI_C }} - 8\pi ^3 e^{ - \alpha /2}
     \frac{{i\Gamma ^4 }}{{p_ -  p_ +  }}\left\{ {S\cos (2\varphi _X  + t_0 )
      + Q\sin (2\varphi _X  + t_0 )} \right\}
\end{equation}
\begin{equation}\label{modulAsym}
    \frac{{\Delta V}}{{RI_C }} = 16\pi ^3 e^{ - \alpha /2} \frac{{i\Gamma ^4 }}
    {{p_ -  p_ +  }}\sqrt {S^2  + Q^2 }
\end{equation}
where $t_0=(i\rho/2\Gamma)\alpha$;
\begin{equation}\label{V0}
    V_0  = \pi RI_C \left( {\frac{\Gamma }{{(1 - \rho )p_ -  }}
    + \frac{\Gamma }{{(1 + \rho )p_ +  }}} \right)
\end{equation}
\begin{equation}\label{P+-}
    p_ \pm   = \int\limits_0^{2\pi } {e^{ - W_ \pm  (y)} dy\int\limits_0^y
    {e^{W_ \pm  (x)} dx}  - \left( {1 - e^{\frac{{i(1 \pm \rho )\pi }}{\Gamma }} }
     \right)^{ - 1} \int\limits_0^{2\pi } {e^{W_ \pm  (x)} dx} \int\limits_0^{2\pi }
      {e^{ - W_ \pm  (x)} dx} }
\end{equation}
\begin{equation}\label{W+-}
    W_ \pm  (x) = \frac{i}{{2\Gamma }}(1 \pm \rho )x + \frac{{1 \pm \gamma }}
    {\Gamma }\cos x
\end{equation}
The quantities $S$ and $Q$ have the forms:
\begin{equation}\label{S}
    S = \frac{i}{{2\Gamma }}\left( {\frac{{1 - \rho }}{{p_ +  }} +
    \frac{{1 + \rho }}{{p_ -  }}} \right)B(\rho ,\gamma )B( - \rho , - \gamma )
\end{equation}
\begin{equation}\label{Q}
   Q = \frac{{B(\rho ,\gamma )A( - \rho , - \gamma )}}{{p_ +  }} -
   \frac{{B( - \rho , - \gamma )A(\rho ,\gamma )}}{{p_ -  }}
\end{equation}
where
\begin{equation}\label{A}
    A(\rho ,\gamma ) = \sum\limits_{n =  - \infty }^{n =  + \infty }
    {\frac{{( - 1)^n nI_n \left( {\frac{{1 + \gamma }}{\Gamma }} \right)I_{n + 1}
    \left( {\frac{{1 + \gamma }}{\Gamma }} \right)}}{{\left( {\frac{i}{2}(1 + \rho )}
     \right)^2  + n^2 \Gamma ^2 }}}
\end{equation}
\begin{equation}\label{B1}
    B(\rho ,\gamma ) = \sum\limits_{n =  - \infty }^{n =  + \infty } {\frac{{( - 1)^n
     I_n \left( {\frac{{1 + \gamma }}{\Gamma }} \right)I_{n + 1} \left( {\frac{{1 +
     \gamma }}{\Gamma }} \right)}}{{\left( {\frac{i}{2}(1 + \rho )} \right)^2  + n^2
     \Gamma ^2 }}}
\end{equation}
For $\alpha\geq1$ the expressions (\ref{VoltAsym})-(\ref{B1}) are
valid at any values of asymmetry parameters $\rho$ and $\gamma$
and at any values of $\beta$  and $\Gamma$, which are consistent
with the condition $\alpha=\pi\beta\Gamma$.
\section{Voltage-Current characteristics}
\subsection{Symmetric DC SQUID}
\begin{figure}[p]
\includegraphics[width=12cm, angle=90]{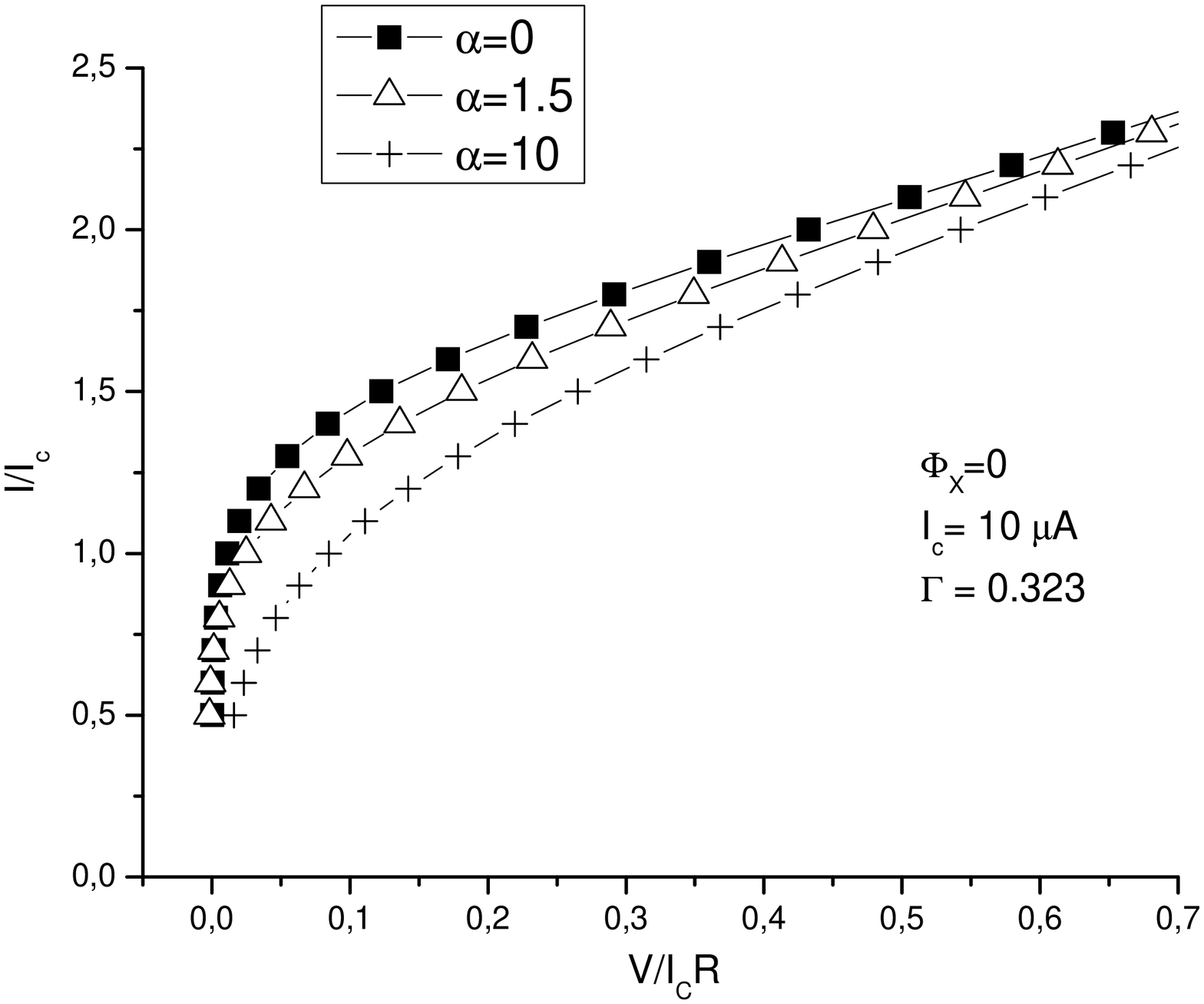} \caption{ VCC of symmetric DC SQUID
at zero magnetic field and different inductances. $I_C=10  \mu A$,
$T=77 K$, the curve with $\alpha=0$ (black box) was calculated
from expression (\ref{voltAmb}), the curves with $\alpha =1.5$
(triangle) and $\alpha=10$ (cross) were calculated from the
expression (\ref{voltsym}).} \label{fig1}
\end{figure}

According to (\ref{voltsym}), the influence of SQUID inductance on
VCC appeared as the reduction of the apparent value of the
critical current (Fig.~\ref{fig1}). Such behavior is due to the
the suppression and masking of the critical current by the
significant thermal current fluctuations in the interferometer
loop. The more the inductance the more the suppression of the
critical current. If $\alpha>>1$ the second term on the righthand
side of the expression (\ref{voltsym}) may be neglected and from
comparison of (\ref{voltsym}) with (\ref{volt}) we see that in
this limit the DC SQUID is equivalent to a single Josephson
junction whose critical current is twice as less as that for the
case $L=0$.

There are two DC SQUID parameters which are easily measured: the
bias current, $I$, and the voltage modulation, $\Delta V$. By the
tuning of the bias current the value $I=I_{MAX}$, which
corresponds to the maximum of the voltage modulation $\Delta
V_{MAX}$, can be found. On the practice this guarantees the
maximum of the SQUID transfer function $dV/d\Phi_X$.

As is seen from (\ref{voltmod}) its righthand side is the product of two terms , one of them $\exp(-\alpha/2)$ depends on the SQUID inductance
only, the other one depends on the bias $I$ and critical $I_C$ currents. The first factor describes the suppression of the critical current  by
the noise current in the interferometer loop. This is similar to the suppression effect in the interferometer loop with a single Josephson
junction \cite{Hlus}. The second factor describes the critical current suppression by the thermal fluctuations, which is similar to the
suppression effect in a single Josephson junction \cite{Amb}. The factorization allows us to carry the first factor to the left hand side of
expression (\ref{voltmod}), so that we consider below the reduced modulation $\Delta V_R=\exp(\alpha/2)\Delta V/R$, which depends on the
critical and bias currents only.

\begin{figure}[p]
\includegraphics[width=12cm, angle=-90]{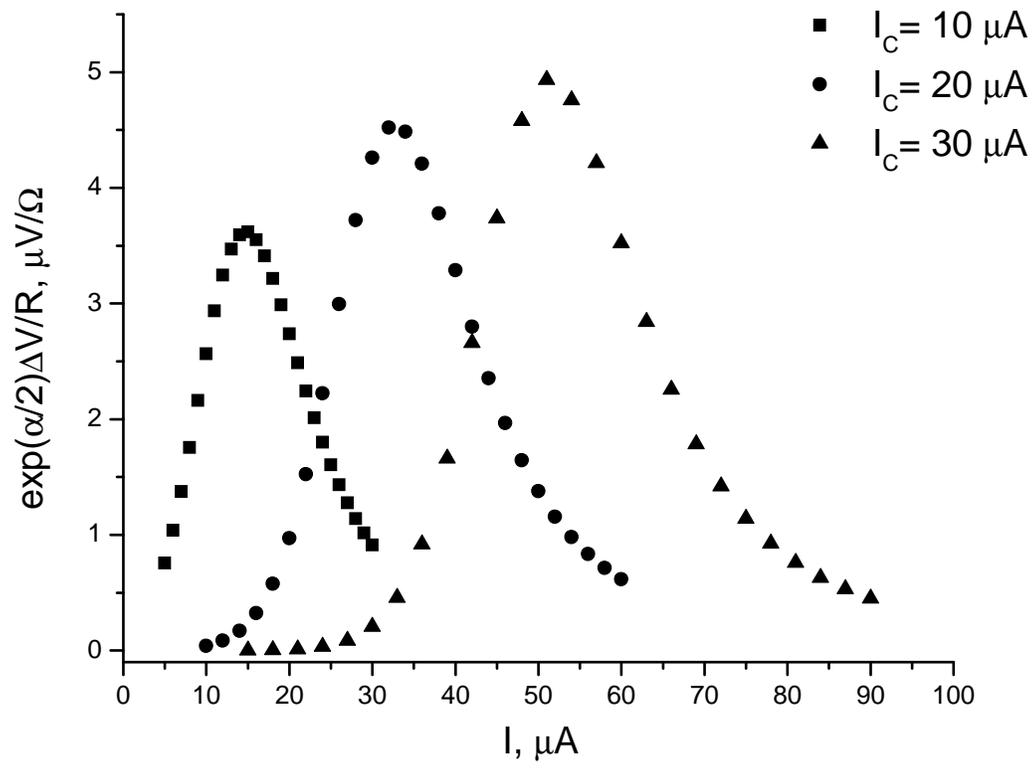} \caption{The reduced voltage
modulation vs bias current curves at different values of the
critical current for symmetric DC SQUID.} \label{fig2}
\end{figure}

The typical dependence of the reduced voltage modulation on the
bias current I at different critical currents is shown in (Fig.
\ref{fig2}). The curves, corresponding to the different $I_C$'s,
are shifted along the current axes and for given $I_C$ each
dependence $\Delta V_R(I)$ has a well defined maximum $\Delta
V_R(I_{MAX})\equiv\Delta V_{R,MAX}$ at the corresponding value of
the bias current.

\begin{figure}[p]
  \includegraphics[width=12cm, angle=-90]{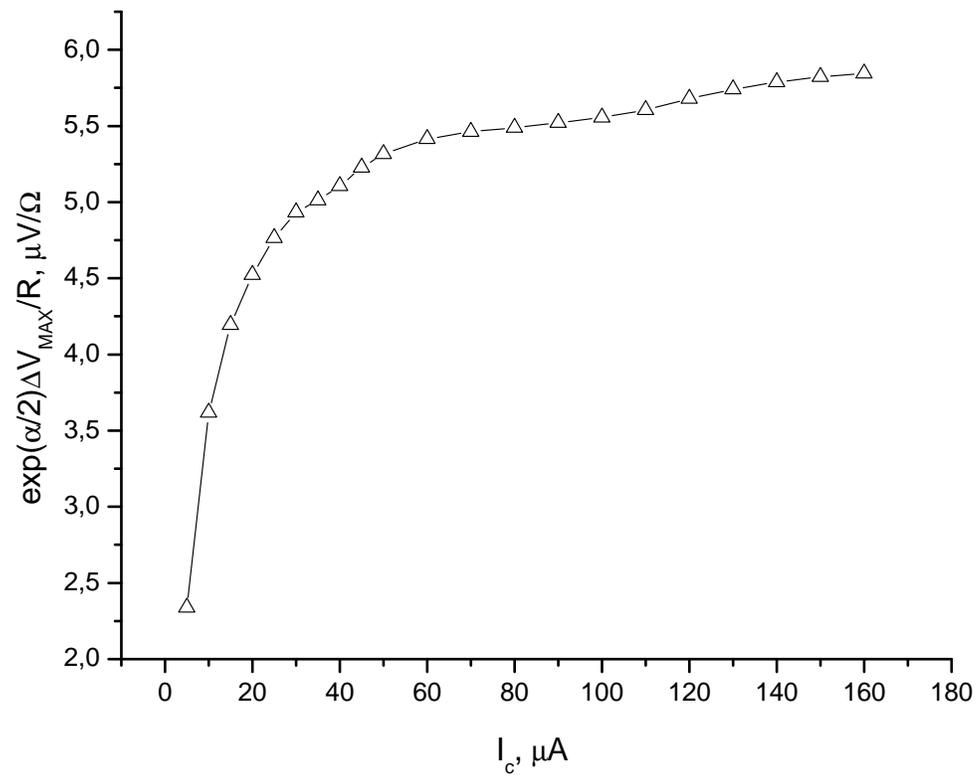}\\
  \caption{The maximum value of the reduced voltage modulation, $\Delta V_{R,MAX}$ vs
  critical current. The different points on the curve correspond to
  the different bias currents $I_{MAX}$, at which $\Delta V$ reaches its
  maximum.}\label{fig3}
\end{figure}

From the equation (\ref{voltmod}) for a set of fixed values of
$I_C$ we have computed a set of the values of maximum voltage
modulation $\Delta V_{R,MAX}$ with the corresponding values of
bias current $I_{MAX}$. In this way we have obtained the table of
values $\Delta V_{R,MAX}(I_C, I_{MAX})$. With the aid of the table
we draw the dependence $\Delta V_{R,MAX}(I_C)$, which is shown in
Fig. \ref{fig3}. It is obvious, that this curve gives the upper
bound of $\Delta V$ for any symmetric DC SQUID. The different
points on the curve correspond to the different bias currents
$I_{MAX}$, at which $\Delta V$ reaches its maximum.

\begin{figure}[p]
  \includegraphics[width=12cm, angle=-90]{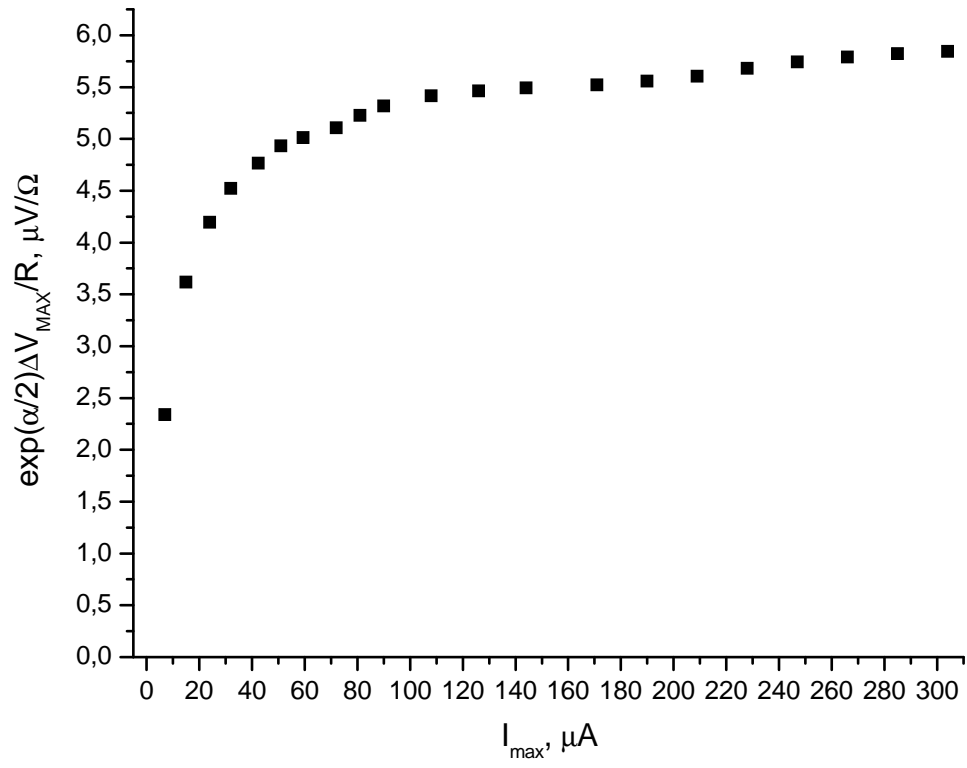}\\
  \caption{Dependence of the maximum value of the reduced voltage modulation
  $\Delta V_{R,MAX}$ on the bias current $I_{MAX}$.}\label{fig4}
\end{figure}

However, as was mention above, the critical current of high-$T_C$
interferometer cannot be measured directly with a good accuracy
because of large thermal fluctuations. Therefore, it is useful in
practice to use the dependence of the maximum modulation signal
$\Delta V_{MAX}$ on the corresponding value of the bias current
$I_{MAX}$. Such dependence, obtained from the table of values
$\Delta V_{R,MAX}(I_C, I_{MAX})$, is shown on Fig. \ref{fig4}. The
different points on this curve belong to the different values of
$I_C$. Every point on the curve is the maximum point on the
corresponding curve from Fig. \ref{fig2}.

The characteristic feature of these two curves $\Delta
V_{R,MAX}(I_C)$ and $\Delta V_{R,MAX}(I_{MAX})$ is the saturation
of $\Delta V_{R,MAX}$ at large critical ($I_C>50 \mu A$) and bias
currents. From the curve shown on Fig. \ref{fig3} the critical
current can be obtained from the measured value of $\Delta
V_{MAX}$. In addition, this dependence allows one to predict the
maximum voltage modulation $\Delta V_{MAX}$, if  the critical
current  of a SQUID is known from the direct measurements.

\begin{figure}[p]
  \includegraphics[width=12cm, angle=-90]{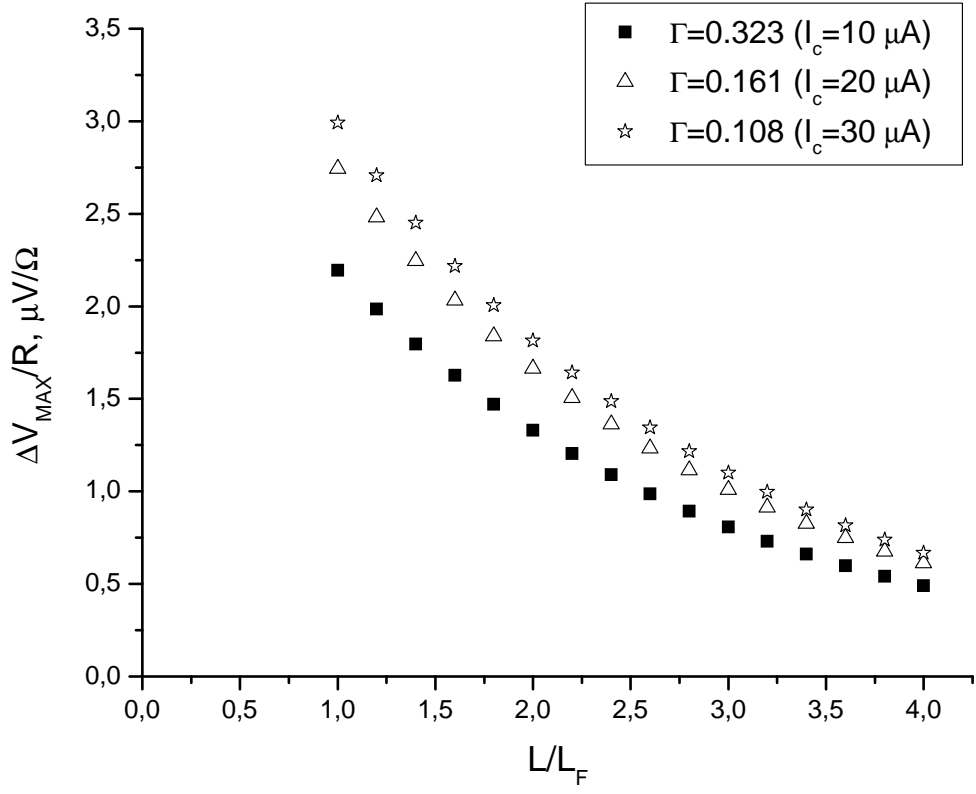}\\
  \caption{The dependence of the maximum value of the voltage modulation
    $\Delta V_{MAX}/R$ on the inductance for symmetric DC SQUID;
    (black box)- $\Gamma=0.323$ $(I_C=10 \mu A)$; (triangle)-$\Gamma=0.161$ ($I_C=20 \mu A$);
     (star)-$\Gamma=0.108$
    ($I_C=30 \mu A$).}\label{fig5}
\end{figure}

The influence of inductance  on $\Delta V_{MAX}$ at different
values of the noise parameter, $\Gamma$ is shown on Fig.
\ref{fig5}. The curves were calculated from Eq.~(\ref{voltmod}).
It can be seen, that the increase of SQUID inductance leads to the
reduction of the voltage modulation in accordance with the scaling
law $\exp(- L/2L_F)$. In addition, the increase of the noise
parameter, $\Gamma$ also leads to the decrease of the voltage
modulation.

\subsection{Critical current}
As is known, it is difficult to measure the critical current of
high-$T_C$ Josephson junctions with a good accuracy because VCC is
washed out by the thermal fluctuations. This problem can be
solved, if we relate the critical current with bias current
$I_{MAX}$, which can be measured directly in SQUID scheme. The
idea was realized in the paper \cite{Drung}, where, based on the
numerical simulations of the paper \cite{Voss}, the following
approximate expression for $I_{MAX}$ has been suggested:

\begin{equation}\label{imax}
    I_{MAX }  = 2I_C \left( {1 - \sqrt {\Gamma /\pi } } \right)
\end{equation}
From Eq.~(\ref{imax}) the critical current can be expressed in
terms of the well measured bias current, $I_{MAX}$:
\begin{equation}\label{ic}
    I_C  = \frac{{I_{MAX } }}{2} + \frac{{k_B T}}{{\Phi _0 }}
    \left( {1 + \sqrt {1 + \frac{{I_{MAX } \Phi _0 }}{{k_B T}}} } \right)
\end{equation}
which for $T=77 K$ becomes:
\begin{equation}\label{ic2}
    I_C=\frac{I_{MAX}}{2}+0.514\left(1+\sqrt{1+1.945I_{MAX}}\right)
\end{equation}
where $I_C$, $I_{MAX}$ are in $\mu A$.

\begin{figure}[p]
  \includegraphics[width=12cm, angle=-90]{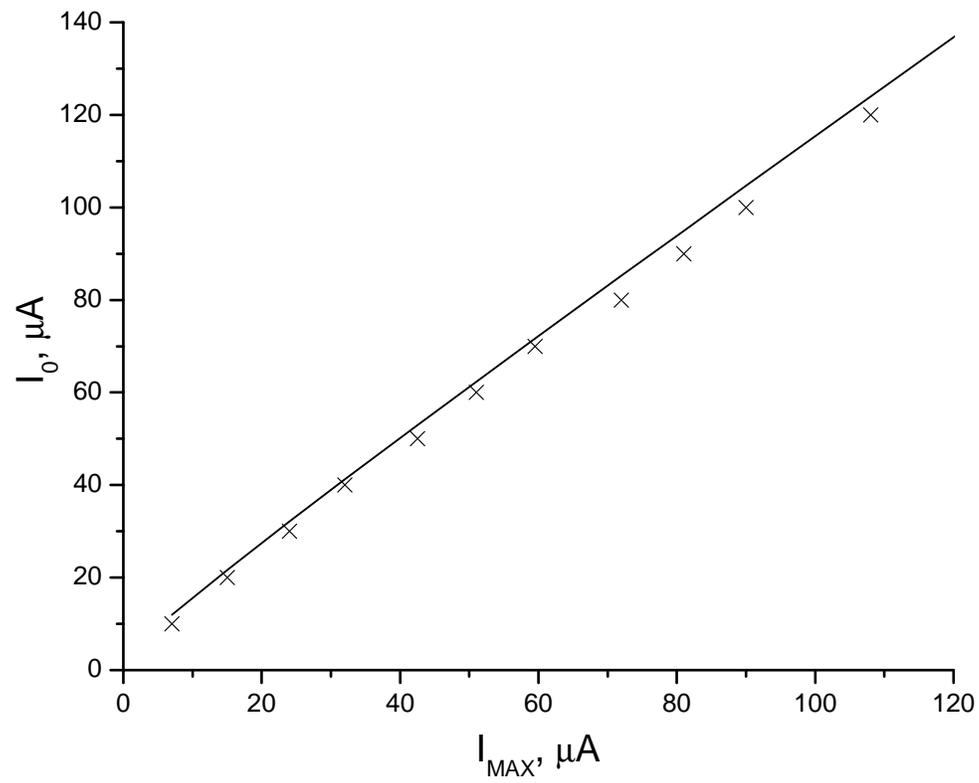}\\
  \caption{The dependence of the SQUID critical current $I_0=2I_C$
  on the bias current $I_{MAX}$. Solid line is the expression of
  Drung (Eq.~(\ref{ic2})); crosses are theoretical points obtained
  from Eq.~(\ref{voltmod}).}\label{fig6}
\end{figure}

The Eq.~(\ref{imax}) has been obtained for low-$T_C$
junctions~\cite{Voss} and applied to high $T_C$ SQUIDs
in~\cite{Drung}. Therefore, it is interesting to compare
~(\ref{ic}) with our theory. As is seen from Fig.~\ref{fig2}, the
expression ~(\ref{voltmod}) establishes a single-valued dependence
between the critical current and the bias current. The bias
current is the well-measured parameter in SQUID scheme. If
$I_{MAX}$ is known from the experiment the expression
~(\ref{voltmod}) permits to find the critical current at which the
voltage modulation reaches its maximum. From ~(\ref{voltmod}) for
a set of critical current values, $I_C$ we have found  a set of
the bias currents $I_{MAX}$, which give the maximum voltage
modulation. The comparison of Drung's expression ~(\ref{ic2}) with
our theory is shown on Fig.~\ref{fig6}. As is seen from the figure
Drung's expression ~(\ref{ic2}) (solid line) and the theory
(crosses) give approximately similar results. The deviation
between Drung's curve and theoretical points is no more than $10
\%$ in the whole range of $I_{MAX}$'s.

\begin{figure}[p]
  \includegraphics[width=12cm, angle=-90]{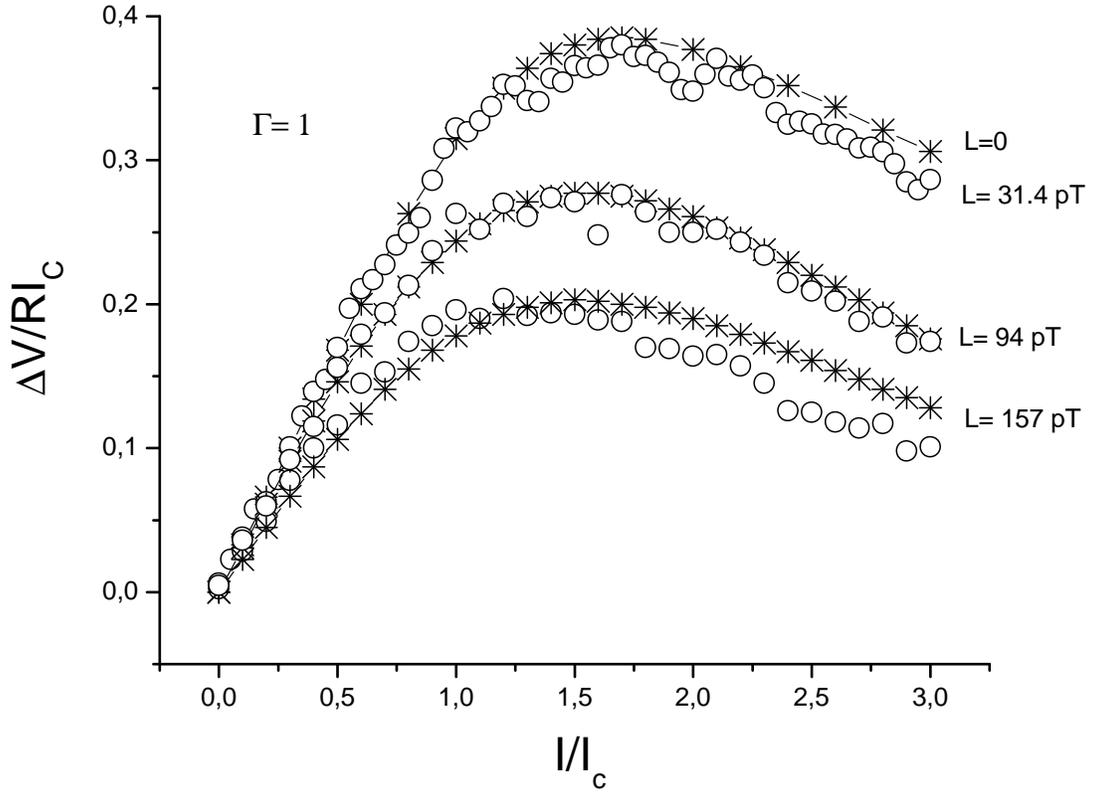}\\
  \caption{The dependence of the normalized voltage modulation on the bias
current. Comparison of theory (stars) with the simulation of
Kleiner (opened circles). The curve for $L=0$ was calculated from
(\ref{voltAmb}) and (\ref{prob}), the curves for $L= 94 pT$,
$L=157 pT$ were calculated from (\ref{voltmod}). All calculations
are made at $77 K$ for $\Gamma=1$.}\label{fig7}
\end{figure}

Below we compare the analytical expression (\ref{voltmod}) with
the results of the computer simulations of the stochastic DC SQUID
equations (\ref{flux}) and (\ref{delta}), obtained by other
authors. The normalized voltage modulation $\Delta V/RI_C$ as a
function of the bias current at the different SQUID inductances is
shown on Fig.~\ref{fig7}. The curves marked by black stars are the
theoretical ones, calculated from Eq.~(\ref{voltmod}) (the curves
for $L=94 pT$, $L=157 pT$) and from the Ambegaokar-Halperin
expressions (\ref{voltAmb}), (\ref{prob}) (the curve for $L=0$).
The curves marked by open circles obtained by computer simulations
of exact stochastic equations (\ref{flux}) and (\ref{delta})
\cite{Kleiner}. All calculations and simulations were made for
$T=77 K$ and noise parameter $\Gamma=1$ ($I_C=3.23 \mu A)$. As is
seen from Fig.~\ref{fig7}, the simulated curves are close to the
calculated ones. Therefore, our expression (\ref{voltmod}) is a
good  approximation of the exact solution of stochastic equations
(\ref{flux}) and (\ref{delta}).

\begin{figure}[p]
  \includegraphics[width=6cm, angle=-90]{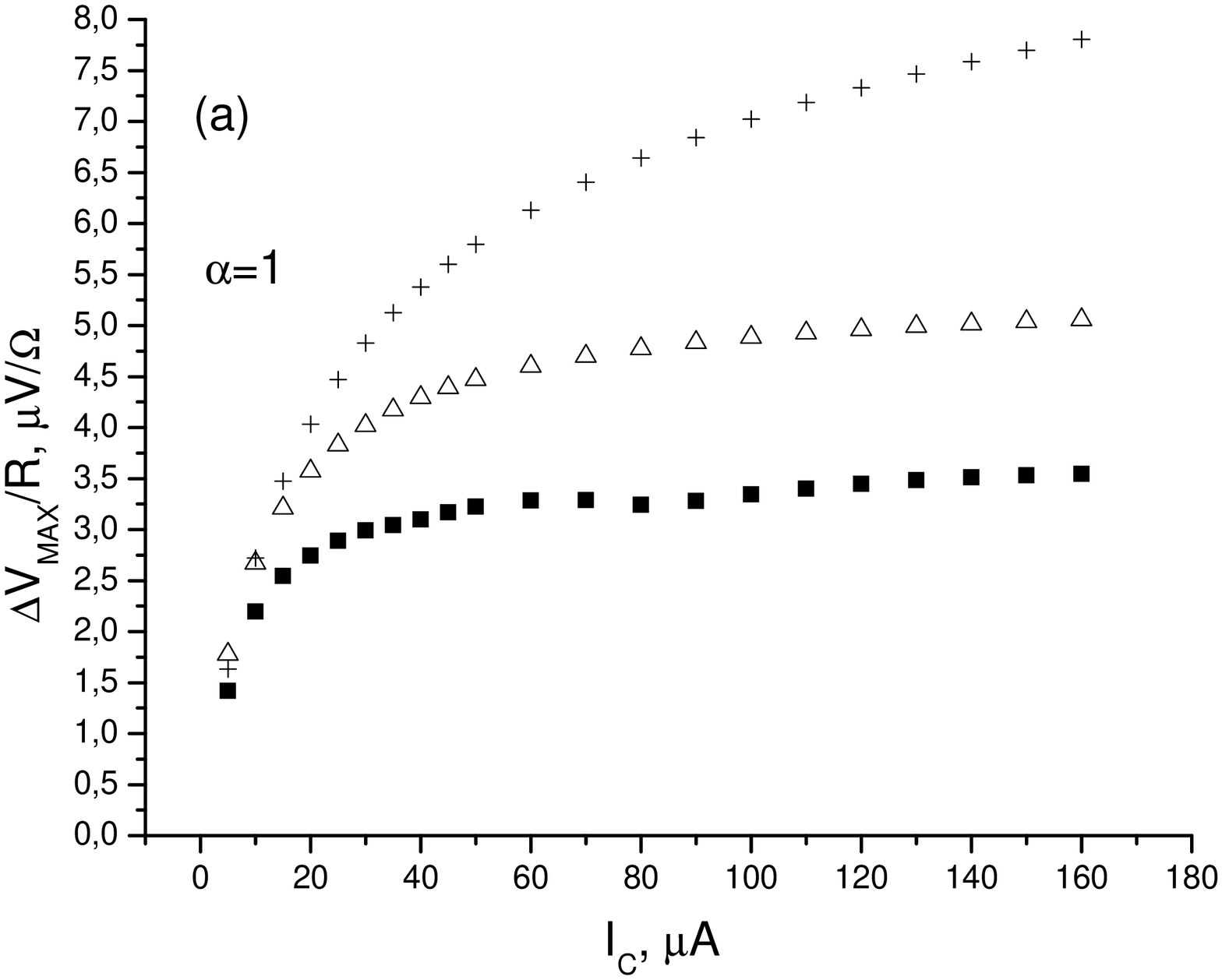}
    \includegraphics[width=6cm, angle=-90]{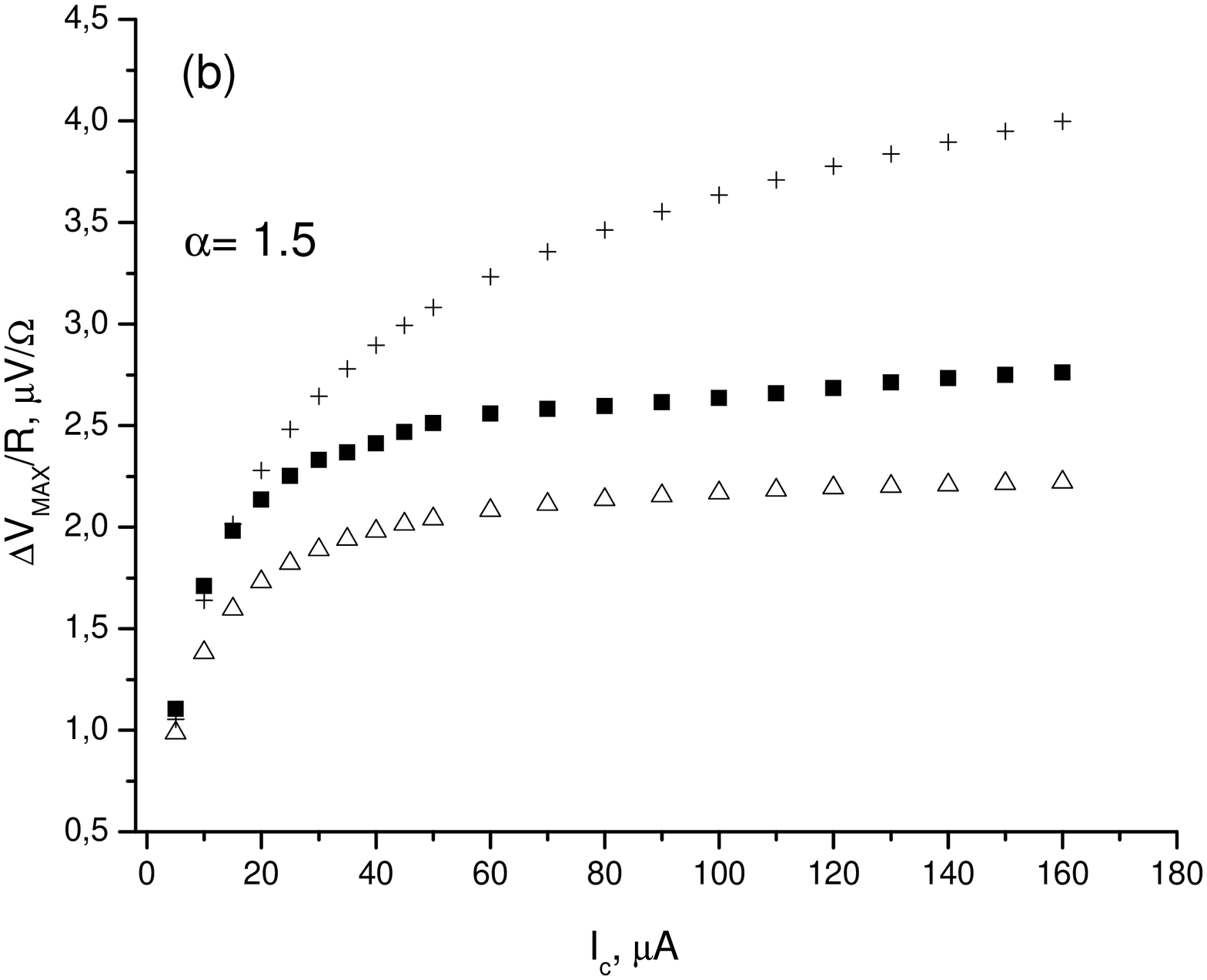}\\
  \includegraphics[width=6cm, angle=-90]{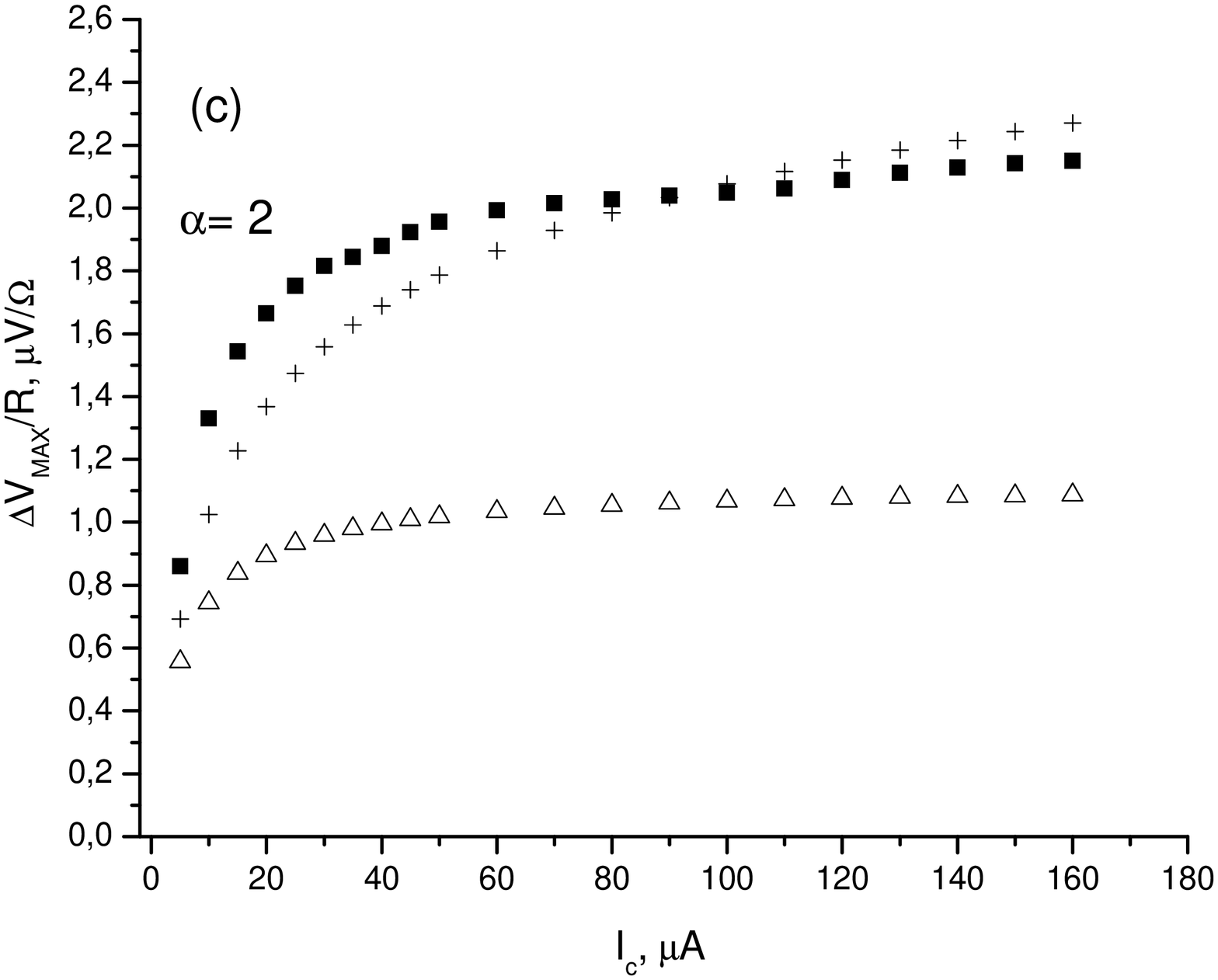}
    \includegraphics[width=6cm, angle=-90]{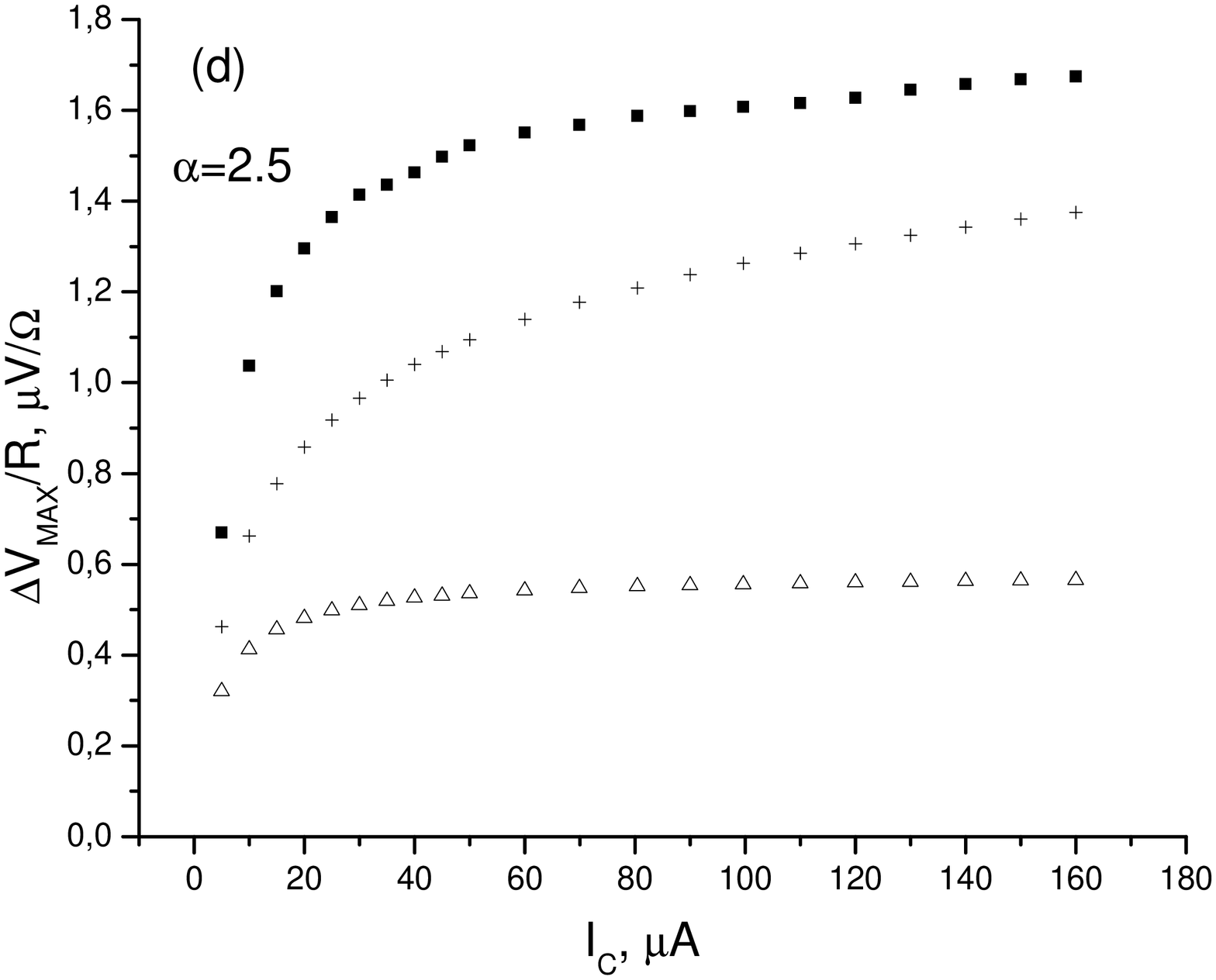}\\
  \caption{The comparison of the maximum voltage modulation given by
    (Eq.~(\ref{voltmod})),(black box) with those obtained from
    the expressions of Kleiner
  (Eq.~(\ref{Kle})),(cross) and Enpuku (Eq.~(\ref{Enp})), (triangle)
  for: (a) $\alpha=1$; (b) $\alpha=1.5$; (c) $\alpha=2$; (d) $\alpha=2.5$.}
  \label{fig8}
\end{figure}

In addition, we compare our theory with the two well known
expressions for the maximum transfer function, $V_\Phi=dV/d\Phi$
which have been obtained by the empirical fit to the computed
values obtained from the simulations of the exact stochastic
equations of DC SQUID. The first expression is obtained in
\cite{Koelle} and is valid for $\beta\geq 0.5$, $\Gamma\leq 1$:
\begin{equation}\label{Kle}
    \frac{{\Delta V_{MAX}}}{{I_C R}} = \frac{{7.3\beta ^{0.15} }}{{\pi (1 + \beta )
    \left[ {\left( {80\Gamma \beta } \right)^{0.4}  + 0.35\left( {4\Gamma \beta }
     \right)^{2.5} } \right]}}.
\end{equation}
The second one is the widely used expression of Enpuku
\cite{Enpuku}:
\begin{equation}\label{Enp}
    \frac{{\Delta V_{MAX}}}{{I_C R}} = \frac{4}{{\pi (1 + \beta )}}\exp
    \left( { - 3.5\pi ^2 \frac{{k_B TL}}{{\Phi _0^2 }}} \right)
\end{equation}
Here the transfer function was recalculated to $\Delta V$ ($\Delta
V=V_\Phi\Phi_0/\pi$) assuming a sine shape of the voltage-to-flux
curve. The comparison of our theory with expressions (\ref{Kle})
and (\ref{Enp}) for several values of the inductances is shown on
Fig.(\ref{fig8}).

It is seen, that the theoretical curve (\ref{voltmod}) and the
curve of Enpuku (\ref{Enp}) reach the saturation at approximately
$I_C>40 \mu A$ while the curve of Kleiner (\ref{Kle}) has a
constant non vanishing slope. This slope is probably due to the
fact that the right hand side of (\ref{Kle}) is equal, in fact, to
the transfer function $V_{\Phi}$ which  has actually been
calculated in \cite{Koelle}. For relatively high critical currents
(small $\Gamma$'s) the sine approximation ($\Delta
V=V_\Phi\Phi_0/\pi$) we made for the shape of the signal is not
very good due to the distortion of the signal shape. It is also
worth noting that the SQUID inductance affects $\Delta V_{MAX}$ in
different ways. The Eq.~(\ref{Kle}) always gives the highest
values, except for $\alpha=2$ for $I_C<80 \mu A$
(Fig.~\ref{fig8}d). The Eq.~(\ref{Enp}) always gives the lowest
values, except for $\alpha=1$ (Fig.~\ref{fig8}a). For large
inductances $\alpha\geq2$ the theoretical curve is always higher.
The influence of inductance is more pronounced for the voltage
modulation given by Eqs.~(\ref{Kle}) and (\ref{Enp}). For example,
for $I_C=100 \mu A$ from $\alpha=1$ to $\alpha=2.5$ the
theoretical voltage modulation is reduced by a factor of two,
while the reduction factor for the Kleiner's and Enpuku's
expressions is six and nine, respectively.
\subsection{Asymmetric DC SQUID}
The numerical simulations of stochastic differential equations,
which govern the dynamics of DC SQUID are very time consuming for
practical high $T_C$ DC SQUIDs due to the large thermal
fluctuations and large loop inductance. This is why in most cases
the investigations of asymmetric DC SQUID are restricted to the
computer simulations in small inductance limit ($\alpha<<1$)
\cite{Testa1, Muller, Enpuk, Testa2, Drung}.  Since our theory is
valid for asymmetric SQUIDs with large inductance we cannot
compare it with the results of other authors obtained for
asymmetric SQUIDs with small inductance.

The practical importance of asymmetric DC SQUIDs is that they have
a higher transfer function (the slope of the voltage-to-flux
curve) as compared to the symmetric case \cite{Testa1, Muller,
Enpuk, Testa2, Drung}. This property is generally attributed to
the distortion of the voltage-to-flux curve. The shape of this
curve significantly differed from the sine shape that results in
the high steep of the slope. For large inductance SQUID the
voltage-to-flux curve has a sine shape as evident from
Eq.~(\ref{VoltAsym}). Therefore, we carefully studied the
influence of the junction asymmetry of large inductance
interferometers on the voltage modulation. We have found that in
the large range of asymmetry parameters the voltage modulation of
asymmetric DC SQUID with large inductance can be significantly
higher than that for symmetric SQUIDs.

\begin{figure}
  \includegraphics[width=12cm, angle=-90]{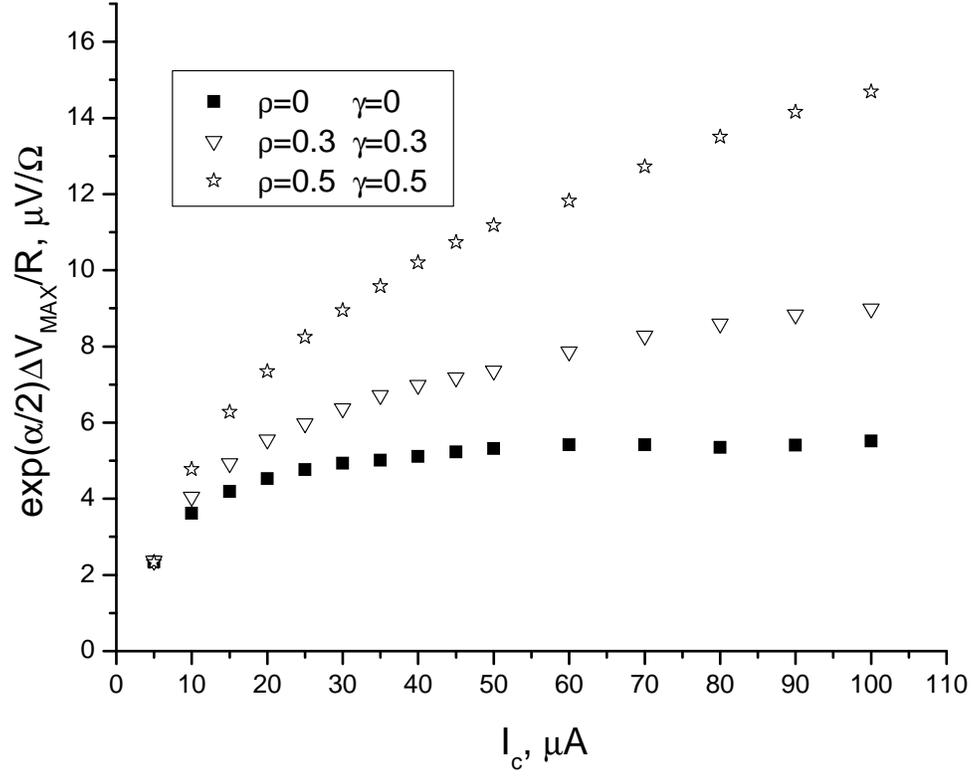}\\
  \caption{The influence of the geometric asymmetry on the
  maximum voltage modulation, $\Delta V_{R,MAX}$.}\label{fig9}
\end{figure}

\begin{figure}
  \includegraphics[width=10cm, angle=-90]{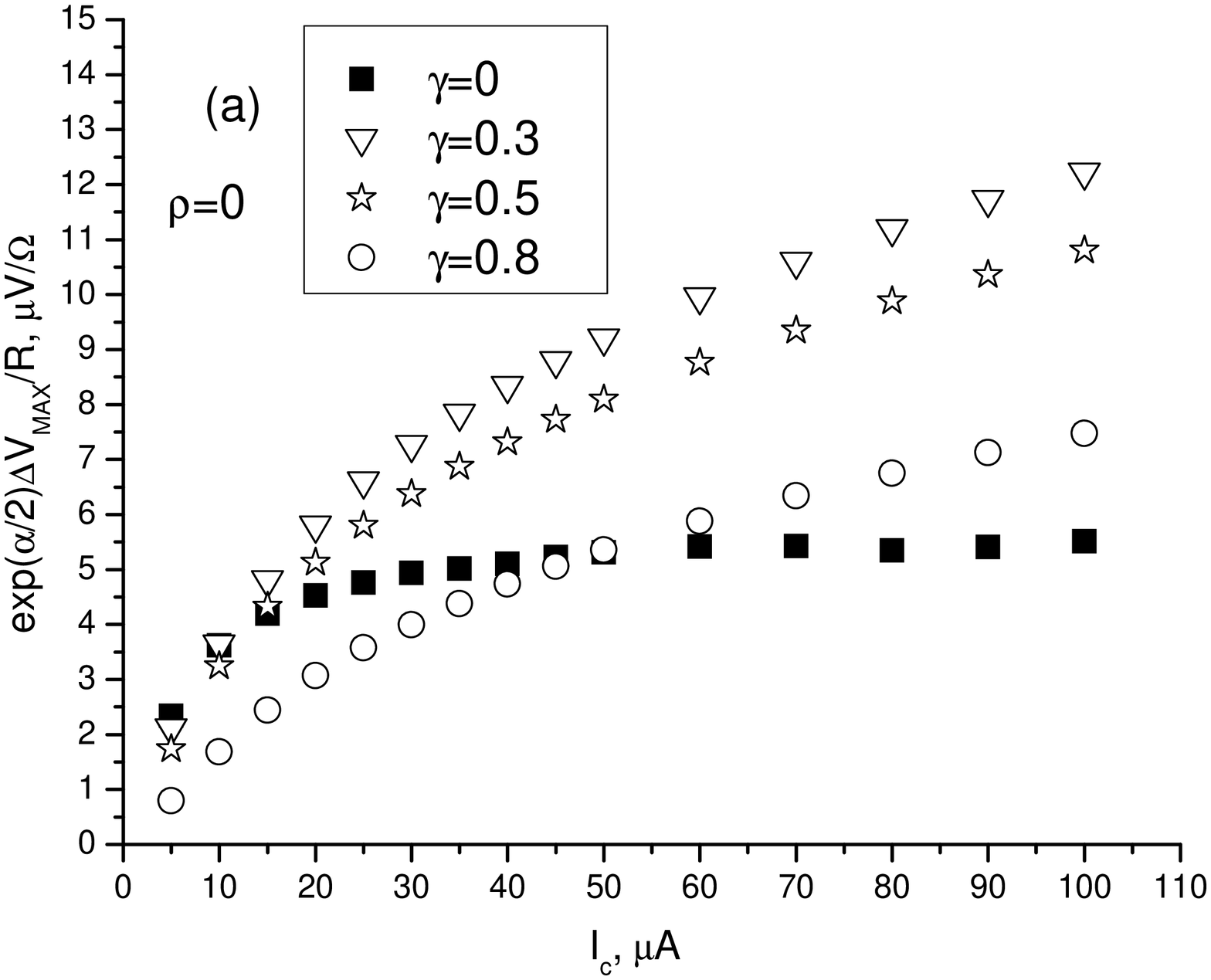}
 \includegraphics[width=10cm, angle=-90]{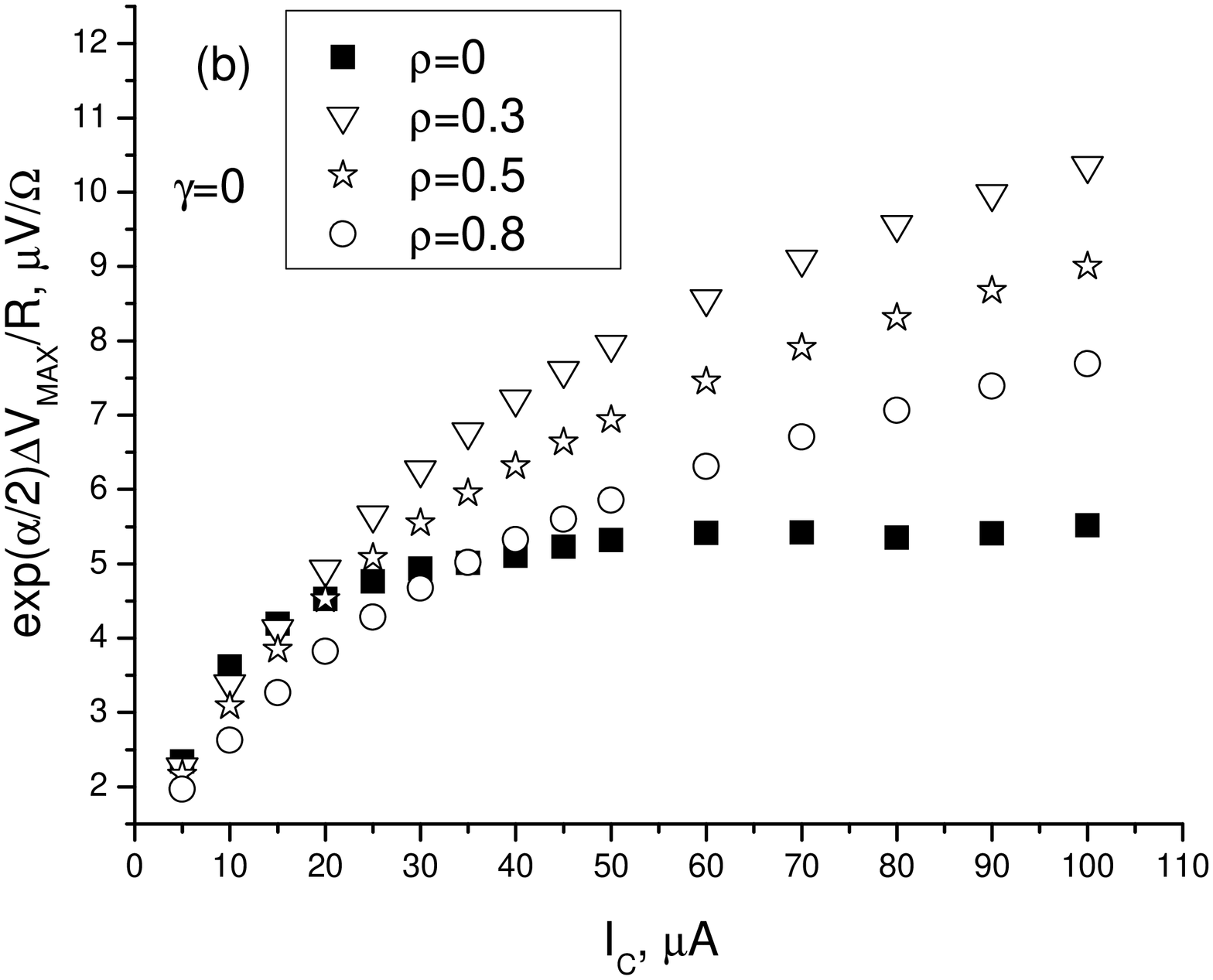}\\
  \caption{The dependence of the maximum voltage modulation, $\Delta V_{R,MAX}$
  on the critical
current for two types of asymmetry: a) the current asymmetry
($\rho=0$); b) the resistance asymmetry ($\gamma=0$)}\label{fig10}
\end{figure}

\begin{figure}
  \includegraphics[width=6cm, angle=-90]{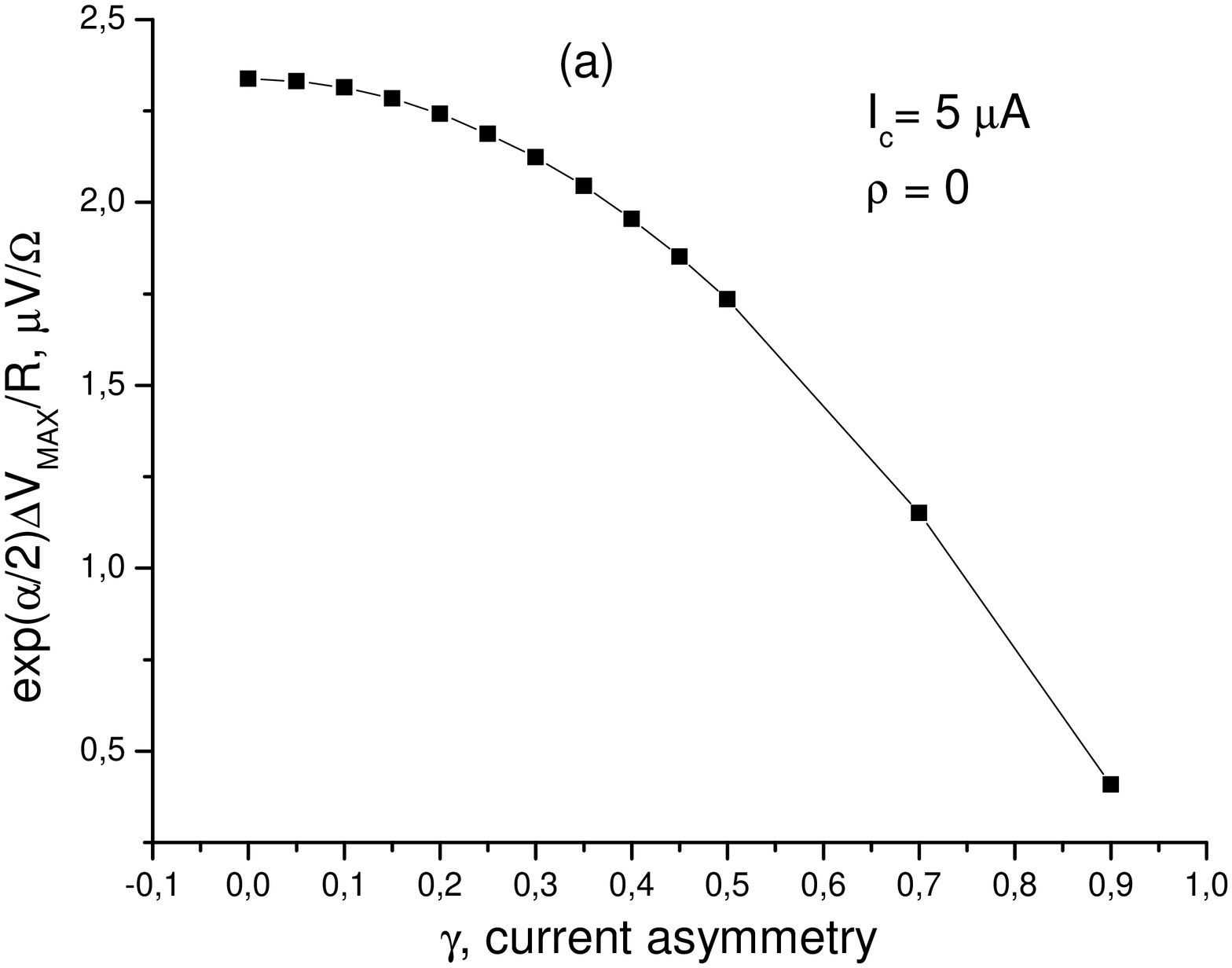}
\includegraphics[width=6cm, angle=-90]{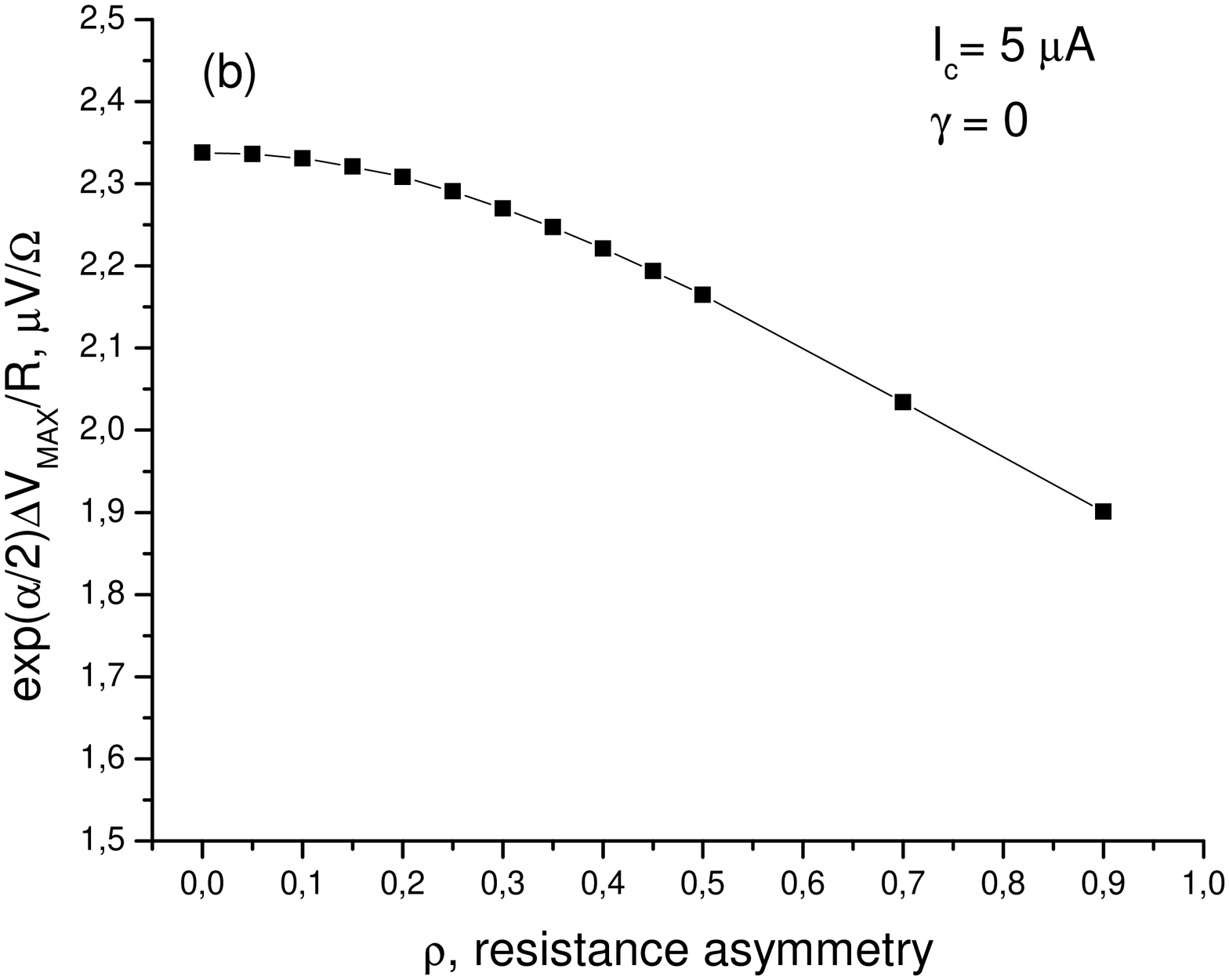}
\includegraphics[width=10cm, angle=-90]{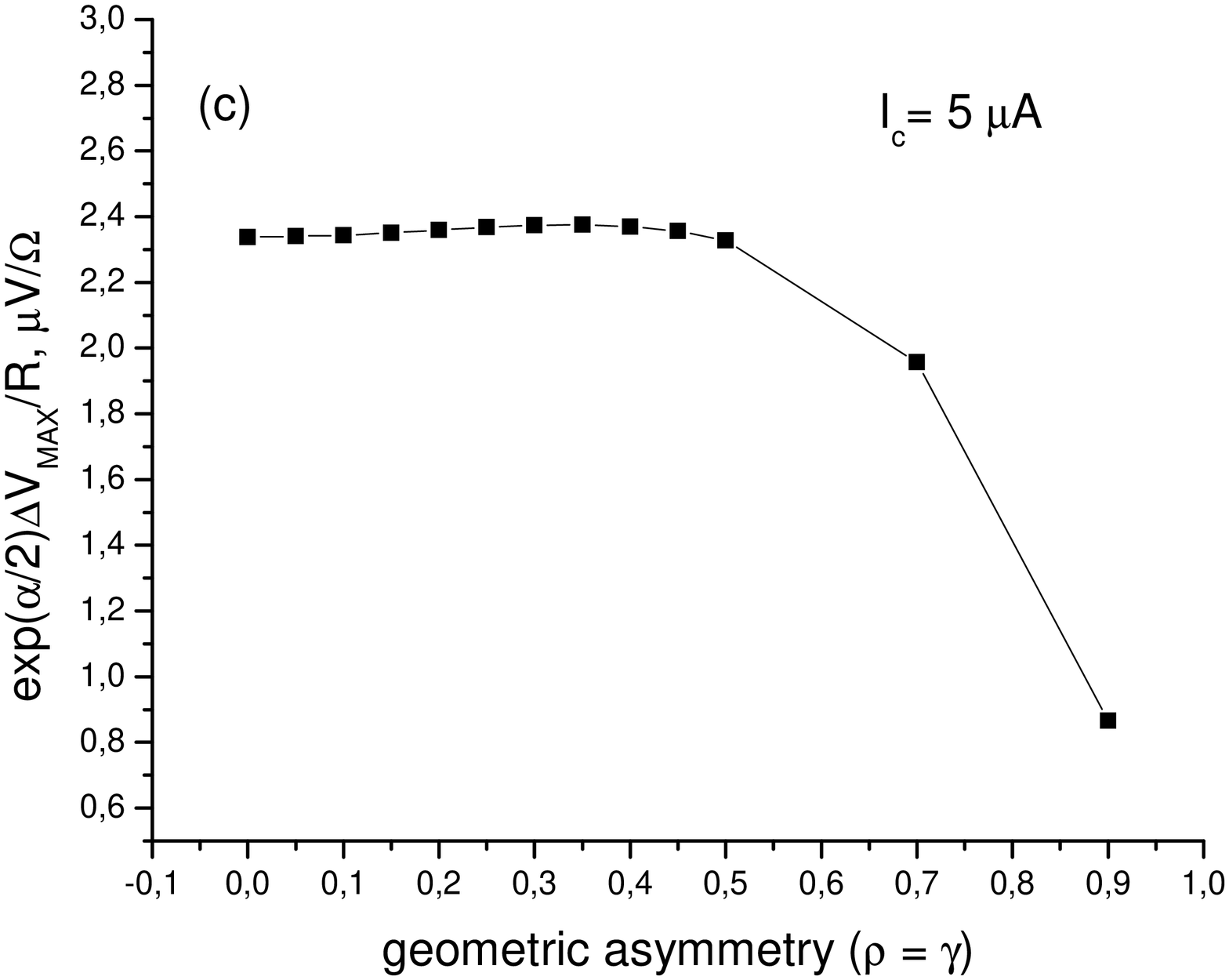}
  \caption{The dependence of reduced maximum voltage modulation on the
  asymmetry parameters  at $I_C=5 \mu A$ for three
types of asymmetry: a) $\rho=0,\gamma\neq0$; b) $\gamma=0,
\rho\neq0$; c) $\rho=\gamma\neq0$.}\label{fig11}
\end{figure}

\begin{figure}
  \includegraphics[width=6cm, angle=-90]{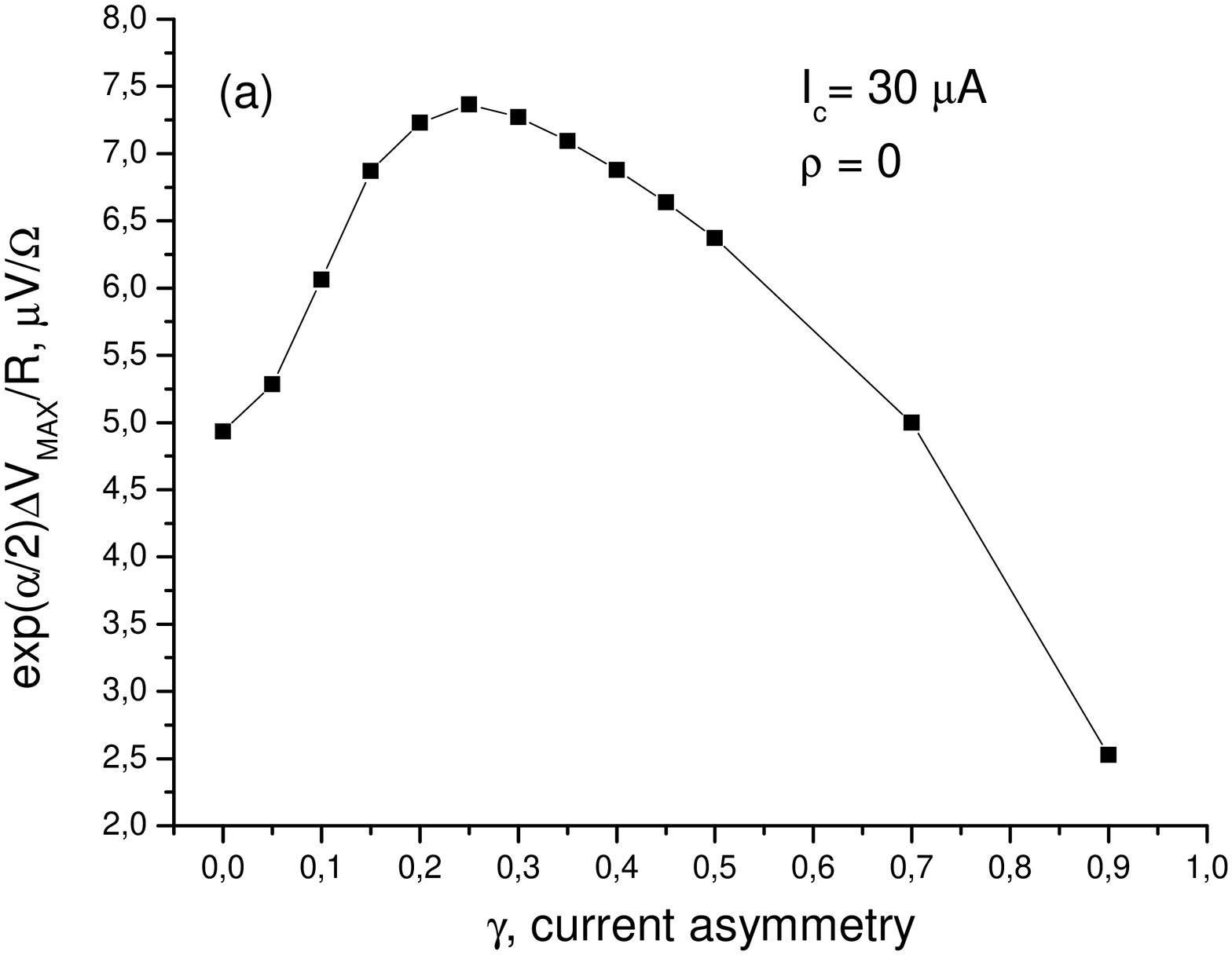}
\includegraphics[width=6cm, angle=-90]{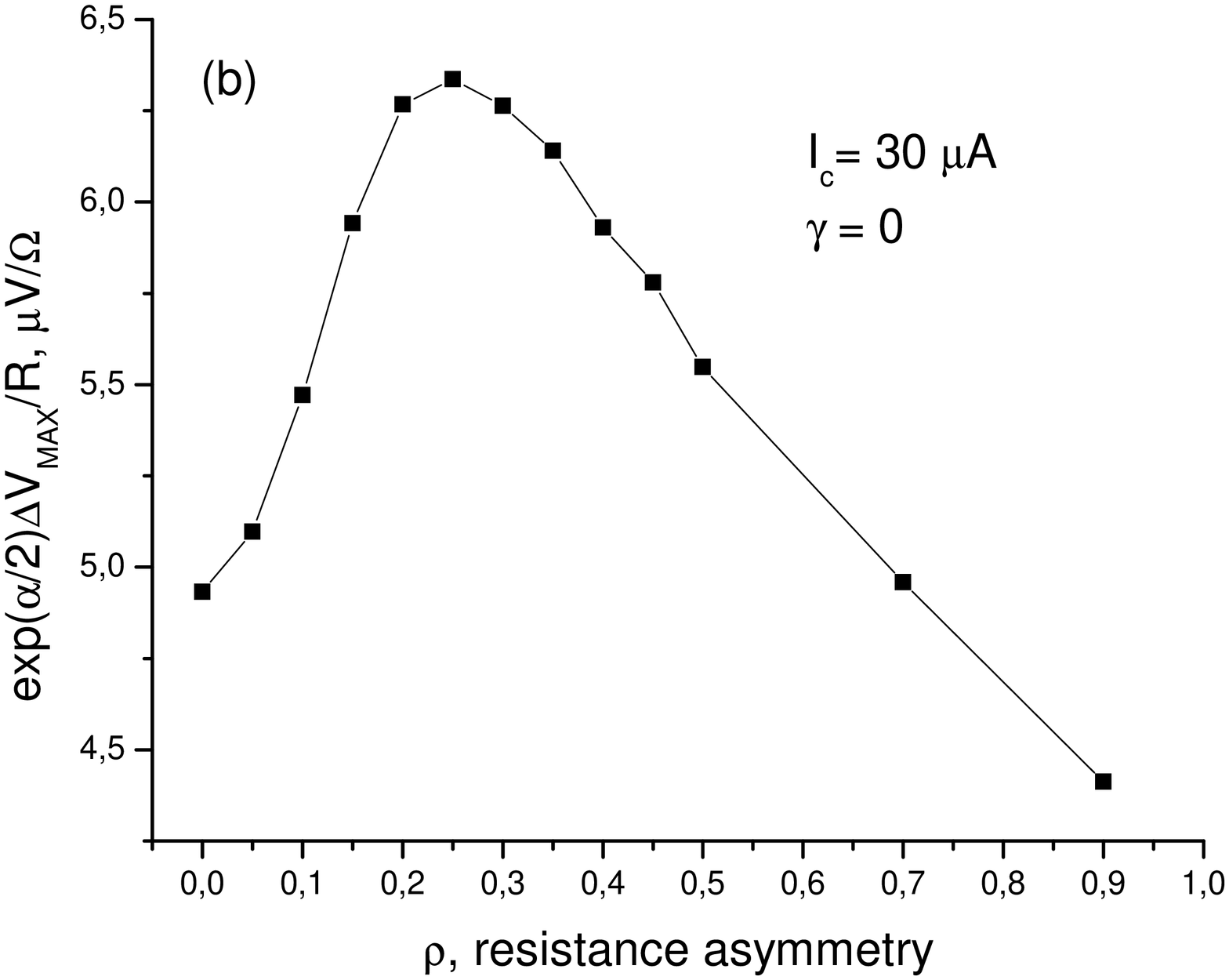}
\includegraphics[width=10cm, angle=-90]{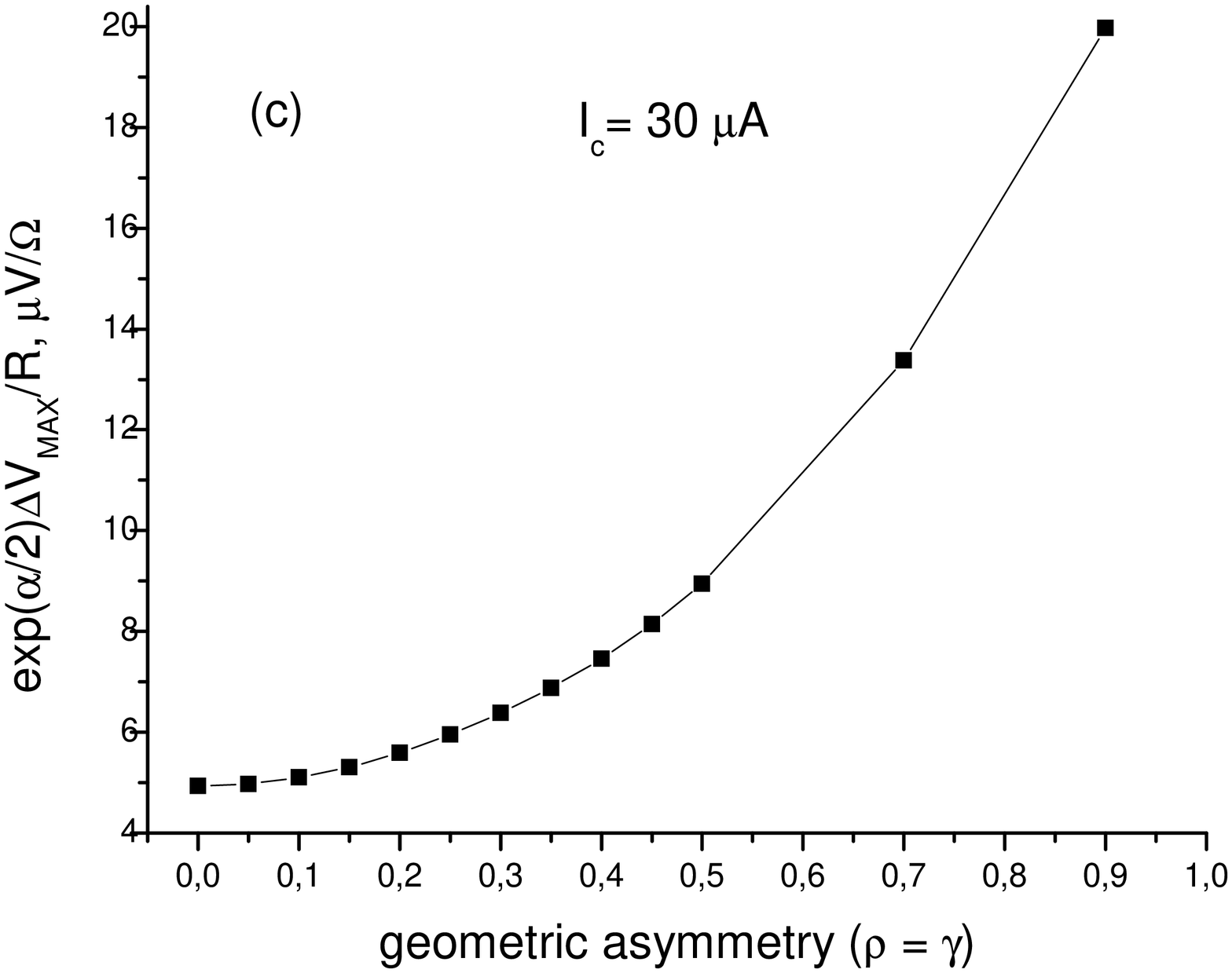}
  \caption{The dependence of reduced maximum voltage modulation on the
  asymmetry parameters  at $I_C=30 \mu A$ for three
types of asymmetry: a) $\rho=0,\gamma\neq0$; b) $\gamma=0,
\rho\neq0$; c) $\rho=\gamma\neq0$.}\label{fig12}
\end{figure}

In principle, the asymmetry parameters $\gamma$ and $\rho$ are
independent of each other and, therefore, they may present in the
junctions in any combination. From our calculations we have chosen
for presentation here only three types of the junction asymmetry
which generally give a correct picture of how the voltage
modulation is influenced by any type of asymmetry. Below we
present the numerical results for the current asymmetry ($\rho=0,
\gamma\neq 0$), the resistance asymmetry ($\rho\neq0, \gamma=0$)
and the geometric asymmetry ($\rho=\gamma\neq 0$) \cite{Muller}.
The effect of different types of the junction asymmetry on the
dependence of the voltage modulation on the critical current is
shown on Fig.~\ref{fig9} and Fig.~\ref{fig10}. As is seen from the
figures, the general trends of the asymmetry are the increase of
the voltage modulation and non vanishing slope of the curves as
compared with the symmetric junctions. The slope being increased
with the increase of the asymmetry. For the geometric asymmetry
the increase of the voltage modulation is seen in the whole range
of the critical currents (Fig.~\ref{fig9}), while for the current
or resistance asymmetry there exists the range of relatively small
critical currents (approximately less than $10 \mu A$) where the
voltage modulation of asymmetric DC SQUID is lower than that of
the symmetric one (Fig.~\ref{fig10}). In order to clarify the
picture we made a careful study of the dependence of $\Delta
V_{R,MAX}$ on the asymmetry parameters at the given values of the
critical current. It appears there exists some threshold value of
the critical current, approximately in the vicinity of $8 \mu A$,
which divides the whole range of the critical currents in two
parts. Below the threshold the current and resistance asymmetry
always leads to the decrease of the voltage modulation as compared
to symmetric DC SQUID (Fig.~\ref{fig11}a, Fig.~\ref{fig11}b). The
geometric asymmetry does not change the voltage modulation up till
approximately $\rho=\gamma=0.5$ with the subsequent decrease of
the voltage modulation (Fig.~\ref{fig11}c). Above the threshold
the curves $\Delta V_{R,MAX}(\gamma, \rho=0, I_C=const)$ and
$\Delta V_{R,MAX}(\gamma=0, \rho, I_C=const)$ have a clear maximum
approximately at the vicinity $\gamma=0.25$ and $\rho=0.25$,
respectively (Fig.~\ref{fig12}a, Fig.~\ref{fig12}b). This maximum
lies approximately 1.5 times higher a symmetric value for the
voltage modulation for current asymmetry and 1.3 times higher  a
symmetric value for the voltage modulation for resistance
asymmetry. It is worth noting that in the wide range of the
current asymmetry $(0<\gamma<0.7)$ or of the resistance asymmetry
$(0<\rho<0.7)$ the voltage modulation is higher than its symmetric
value. However, the geometric asymmetry leads to the significant
increase of the voltage modulation in the whole range of the
critical currents (Fig.~\ref{fig12}c).
\section{Comparison with experiment}
In order to compare our theory with experiment we took two groups of DC SQUIDs. The first group comprised about 50 SQUIDs which have been chosen
before for the same purpose \cite{Green3}, \cite{Green4}. All of these SQUIDs are single layer ones, using 100 or 200 nm thick
$YBa_2Cu_3O_{7-x}$ films deposited by laser ablation onto $SrTiO_3$ bicrystal substrates with $24^\circ$ or $30^\circ$ misorientation angles,
both having symmetrical configuration. The technology is described in detail in \cite{Ijss}. The second group comprised 10 bicrystal SQUIDs,
which were also single layer ones, using 150 - 200 nm thick $YBa_2Cu_3O_{7-x}$ films deposited by linear hollow cathode sputtering onto
$SrTiO_3$ bicrystal symmetrical substrates with $24^\circ$ misorientation angle \cite{Lind}. All SQUIDs have $\alpha\geq1$ and $\Gamma\geq0.05$
and all measurements were performed in liquid nitrogen at 77 K.

\begin{figure}
  \includegraphics[width=12cm, angle=-90]{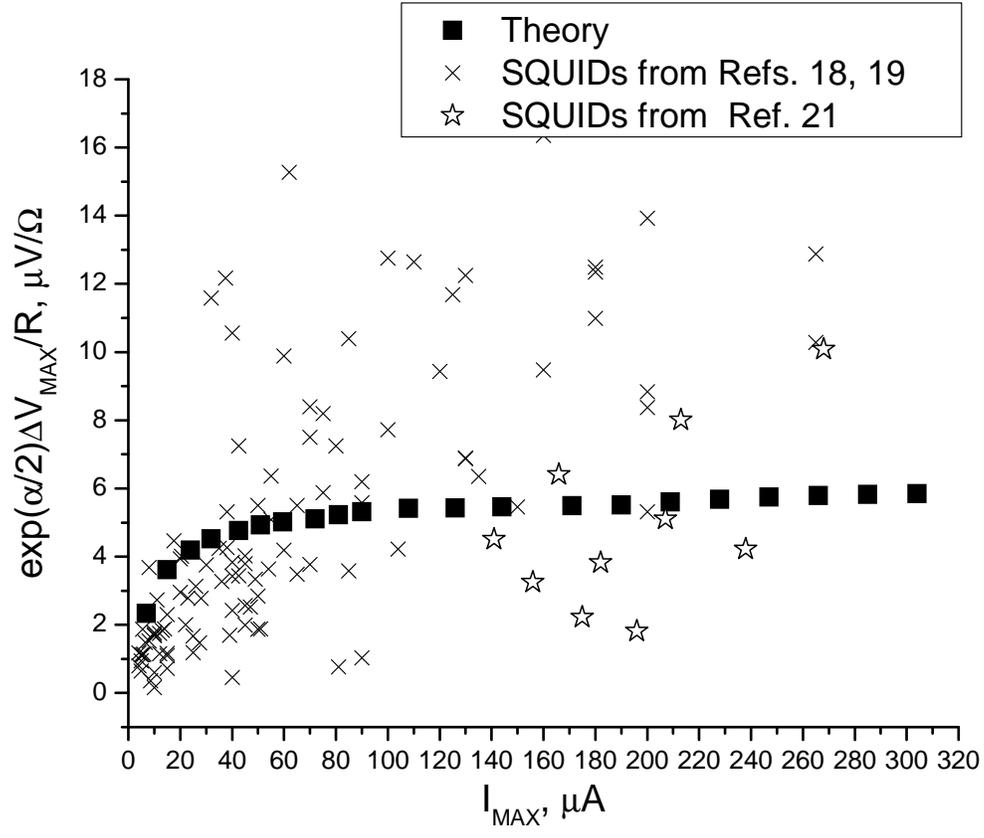}\\
  \caption{The dependence $\Delta V_{R,MAX}(I_{MAX})$ for symmetric
SQUID together with experimental points.}\label{fig13}
\end{figure}

\begin{figure}
  \includegraphics[width=10cm, angle=-90]{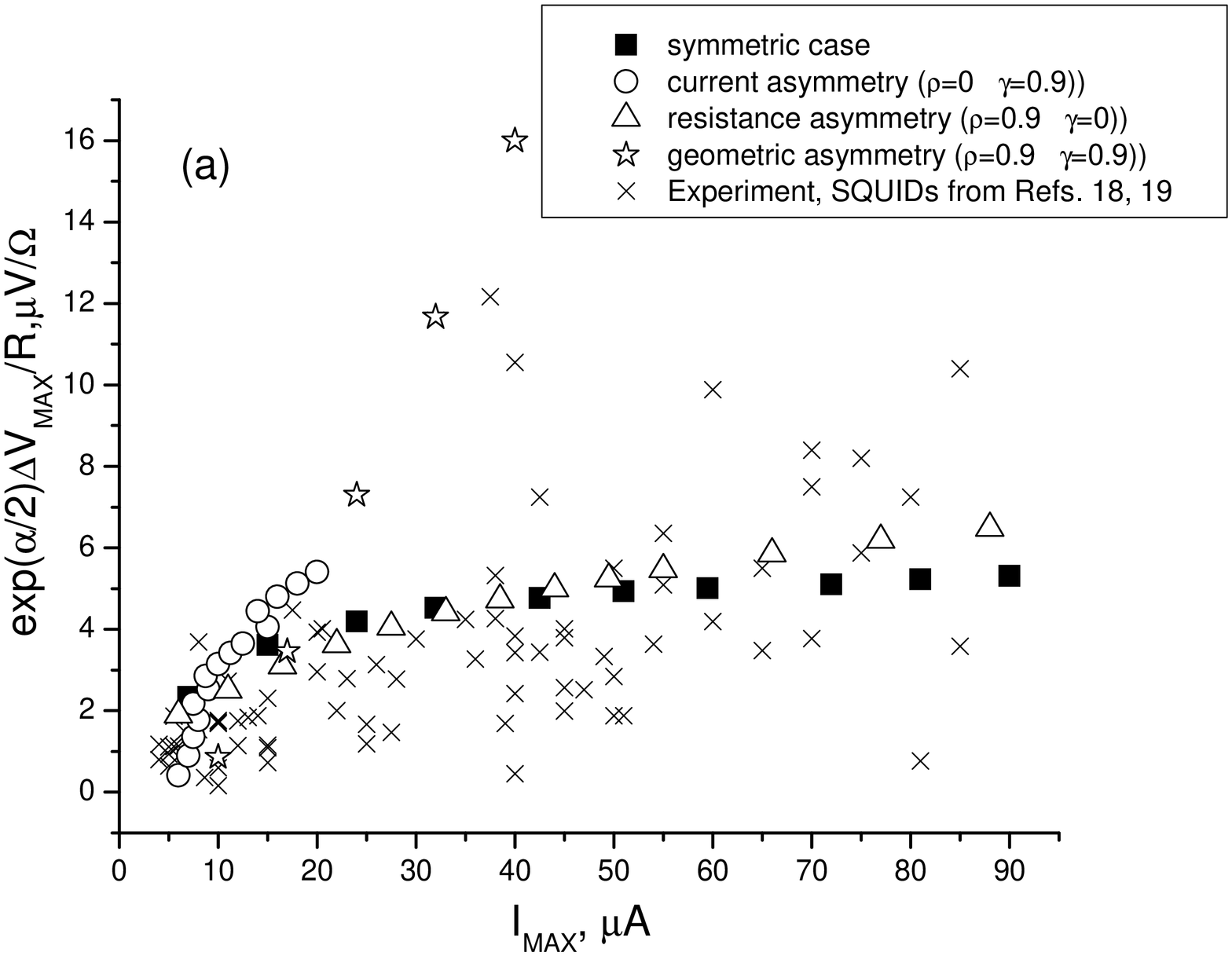}
\includegraphics[width=10cm, angle=-90]{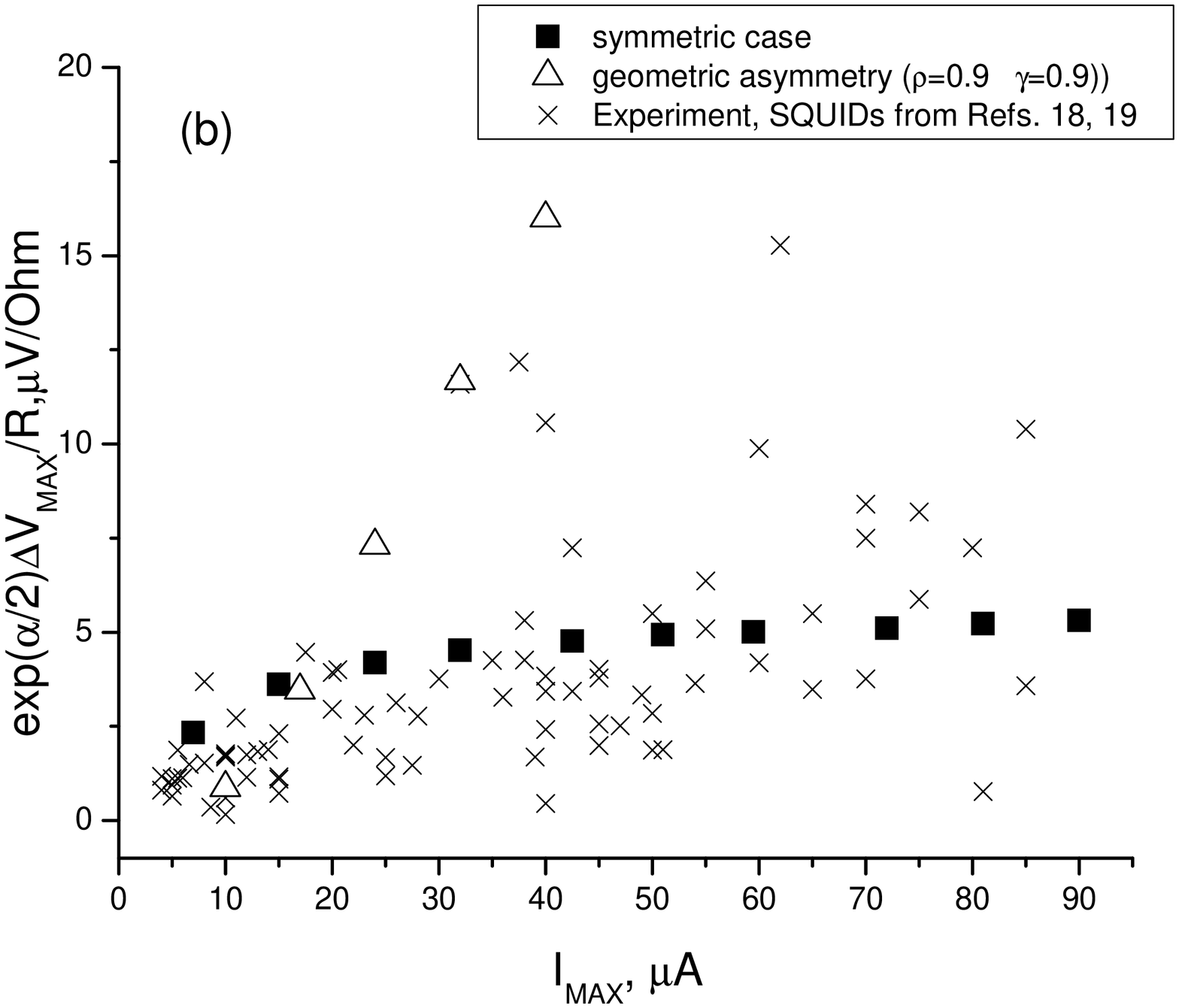}\\
  \caption{The dependence $\Delta V_{R,MAX}(I_{MAX})$ for symmetric
  and asymmetric SQUIDs together with experimental points.}\label{fig14}
\end{figure}

Since the critical current, $I_C$ of a high $T_C$ DC SQUID cannot
be measured with a good accuracy due to high level of thermal
fluctuations, we here, for the comparison with experiment, use
only the quantities, which are measured directly, $\Delta V_{MAX}$
and $I_{MAX}$. The dependence $\Delta V_{R,MAX}(I_{MAX})$ for
symmetric SQUID together with experimental points is shown on
Fig.~\ref{fig13}. It can be seen, that there is a significant
deviation of experimental points from theoretical line of
symmetric DC SQUIDs. Part of the experimental points which lie
above theoretical line can be explained by the junction asymmetry.
In order to show this we add to the plot of Fig.~\ref{fig13} the
theoretical curves for asymmetric DC SQUID (Fig.~\ref{fig14}a and
Fig.~\ref{fig14}b). We choose a significant asymmetry on these
graphs ($\gamma=0.9; \rho=0.9$) in order to mark the borders of
the possible scattering of the voltage modulation values. In
addition, the asymmetry allows one to explain the experimental
points which lie below the symmetric line at small bias currents
$I_{MAX}<10 \mu A$. Since at small bias currents $I_{MAX}$ is
close to $I_C$ (see Eq.~(\ref{ic2})) this reduction of the voltage
modulation is consistent with the result of section IIIC where we
showed that at relatively small critical currents a junction
asymmetry reduced the voltage modulation. Therefore, the junction
asymmetry can explain the experimental values of the voltage
modulation which lie above the line for symmetric DC SQUID and the
points which lie below symmetric line at small bias currents.
However, a significant reduction of the voltage modulation, which
lies well below a symmetric line in Figs.~\ref{fig13} and
\ref{fig14} cannot be explained by the junction asymmetry. A
possible explanation of these points is the presence of relatively
large amplitude of a second harmonic in the junction current phase
relation \cite{Green4}.
\section{Conclusion}
In the paper we consistently compared the analytical theory for
the voltage-current characteristics of the large inductance
($L>100$pH) high-$T_C$ DC SQUIDs that has been developed
previously \cite{Green1, Green2} with the computer simulations and
the experiment. It is shown that the theoretical voltage
modulation for symmetric junctions is in a good agreement with the
results of known computer simulations. It is also shown that the
asymmetry of the junctions results in the increase of the voltage
modulation in the large range of critical currents and asymmetry
parameters. We compared our theory with the experimental values of
the voltage modulation. It appeared  that the asymmetry can
explain a large portion of experimental values of the voltage
modulation which lie above the theoretical curve for symmetric DC
SQUID. It also explains experimental points which lie below the
curve at small critical currents. However, a significant portion
of these values which lie below the curve cannot be explained by
the junction asymmetry. From our opinion a possible explanation of
these low lying points is the presence of relatively large
amplitude of a second harmonic in the junction current phase
relation \cite{Green4}.

\begin{acknowledgments}
We thank R. Kleiner  for providing us with the unpublished results of the numerical simulations shown in Fig. 7 of the paper. We are also
grateful to V. Schultze and R. Kleiner for fruitful discussion.

The authors acknowledge the support by the INTAS grant 2001-0809.
\end{acknowledgments}




\end{document}